\documentclass[a4paper,useAMS,usenatbib]{mnras}
\usepackage{amsmath}
\usepackage{amssymb}
\usepackage{float}
\usepackage{graphicx}
\usepackage{subcaption}
\usepackage{mathptmx}
\usepackage{newtxtext,newtxmath}

\newcommand{\msun}{M$_{\odot}$}
\newcommand{\mseed}[1]{$M_{\mathrm{seed}}=10^{#1}\,\mathrm{M}_\odot$}
\newcommand{\ramses}{\textsc{ramses}}
\newcommand{\vect}[1]{\bmath{#1}}
\newcommand{\upright}[1]{\mathrm{#1}}
\newcommand{\unit}[1]{\, \mathrm{#1}}
\newcommand{\sub}[1]{_{\mathrm{#1}}}

\graphicspath{{./figures_dynamics/}}

\def\equationautorefname~#1\null{Eq.~(#1)\null}
\title[SMBHs dynamics in gas-rich galaxies]{On the Dynamics of Supermassive Black Holes in Gas-Rich, Star-Forming Galaxies: the Case for Nuclear Star Cluster Coevolution} 

\author[P. Biernacki et al.] {Pawel Biernacki\thanks{biernack@physik.uzh.ch}$$, Romain Teyssier$$ and Andreas Bleuler$$\\ \\
  {Center for Theoretical Astrophysics and Cosmology, Institute for Computational Science, University of Zurich,} \\ 
  {Winterthurerstrasse 190, 8057 Zurich, Switzerland} \\ 
}

\begin{document}
\maketitle
\begin{abstract}
We introduce a new model for the formation and evolution of supermassive black holes (SMBHs) in the \ramses{} code using sink particles, 
improving over previous work the treatment of gas accretion and dynamical evolution. 
This new model is tested against a suite of high-resolution simulations of an isolated, gas-rich, cooling halo. 
We study the effect of various feedback models on the SMBH growth and its dynamics within the galaxy. In runs without any feedback, the SMBH is trapped within a massive bulge and is therefore able to grow quickly, 
but only if the seed mass is chosen larger than the minimum Jeans mass resolved by the simulation. 
We demonstrate that, in the absence of supernovae (SN) feedback, the maximum SMBH mass is reached 
when Active Galactic Nucleus (AGN) heating balances gas cooling in the nuclear region. When our efficient SN feedback is included, it completely prevents bulge formation, so that massive gas clumps can perturb the SMBH orbit, and reduce the accretion rate significantly. 
To overcome this issue, we propose an observationally motivated model for the joint evolution of the SMBH and a parent nuclear star cluster (NSC),
which allows the SMBH to remain in the nuclear region, grow fast and resist external perturbations. 
In this scenario, however, SN feedback controls the gas supply and the maximum SMBH mass now depends on the balance between AGN heating and gravity. 
We conclude that SMBH/NSC co-evolution is crucial for the growth of SMBH in high-z galaxies, the progenitors of massive ellipticals today.
\end{abstract}

\begin{keywords}
methods: numerical - galaxies: evolution - galaxies: active - quasars: supermassive black holes - galaxies: star clusters: general
\end{keywords}

\section{Introduction}\label{sec:intro}

Supermassive Black Holes (SMBH) are found in the central region of massive galaxies at all redshifts, mostly in the form of Active Galactic Nuclei (AGN). There is accumulating evidence that SMBH are tightly linked to the evolution of their host galaxy  \citep{Richstone1998, Ferrarese2000, Gebhardt2000, Marconi2003, Haring2004, Kormendy2013}, putting AGN physics at the centre of our understanding of galaxy evolution.  The strong correlation of SMBH masses and stellar velocity dispersion, for example, suggests a possible co-evolution of the central SMBH and its host galaxy \citep{Magorrian1998, Laor2001, McLure2002, Haring2004}. AGN feedback is also often invoked as one of the possible origins of the quenching of star formation in elliptical galaxies \citep{Schawinski2007, Nandra2007, Fabian2012, Yesuf2014, Cheung2016}. The formation of the SMBH themselves remains a mystery. Two main scenarios are considered leading to massive enough SMBH: 1) direct collapse of massive clumps of pristine gas \citep{Loeb1994,Bromm2003} or 2) mergers of stellar remnants in dense stellar clusters \citep{Quinlan1990, PortegiesZwart1999, Devecchi2009}, each scenario having clear strengths and weaknesses, as explained in the reviews of \cite{Begelman2006} and \cite{Volonteri2010}. 
	
Motivated by these observational hints, theoretical models of SMBH growth and their associated feedback (mostly based on complex numerical simulations)  became in recent years more and more sophisticated, with mixed successes when compared against observational data \citep{Springel2005, DiMatteo2005, Bower2006, Croton2006, Hopkins2006, Ciotti2010, Teyssier2011, Dubois2011}. AGN feedback in theoretical models of galaxy formation has proven very efficient at regulating the Star Formation Rate (SFR) in massive, red and dead galaxies, but the X-ray properties of the intergalactic gas are very difficult to reproduce. One natural explanation to the difficulties of these models is the formidable range of scales one has to capture, in order to resolve numerically the entire accretion flow from parsec scales towards the last stable orbit (typically $10^{-5}$~pc). Numerical implementation of SMBH formation, their accretion flows and associated energetic outflows, have to rely on strong approximations, usually referred to as ``subgrid models". Note that the same technique is applied to star formation recipes in galaxy formation simulations, making the whole endeavour of simulating galaxies very challenging. 

As the resolution of galaxy formation simulations is increasing, from thousands of pc in large-scale cosmological simulations \citep{Dubois2014c, Vogelsberger2014, Schaye2015, Dubois2016}, to hundreds of pc in cosmological zoom-in simulations of galaxy formation \citep[e.g.][]{Kim2011, Angles-Alcazar2014, Dubois2015}, ultimately reaching a few pc in isolated discs simulations \citep{Gabor2013, Hopkins2014}, these subgrid models need to be tuned and adapted to the increasingly better resolved interstellar medium (ISM) structure, with an increasingly stronger supersonic turbulence.

The goal of this paper is precisely to study such a model of SMBH formation, growth and feedback in highly resolved, turbulent and clumpy galactic discs, typical of high redshift, gas-rich galaxies \citep{Elmegreen2008a, Dekel2009, Bournaud2012}. This environment is particularly relevant to SMBH physics, as these clumpy discs are believed to be the progenitors of the giant ellipticals hosting the most massive SMBHs in our present epoch \citep{Kormendy2013,McConnell2013}. 

Numerical models of SMBH formation and evolution are all based on the so-called ``sink particle" technique. The SMBH is represented by a point mass, moving through the fluid and interacting with it through accretion and ejection of mass, energy and momentum. Sink particles were first implemented in simulations of star-forming turbulent molecular clouds \citep{Bate1995}, using a Smoothed Particle Hydrodynamics (SPH) code. \citet{Krumholz2004}  was the first one to propose a sink particle implementation for grid-based codes, using Adaptive Mesh Refinement (AMR). The sink particle technique was then adapted to the SMBH formation and evolution, here again first in SPH codes \citep{Springel2005, DiMatteo2005} and then later in AMR codes \citep{Dubois2010, Kim2011}. The key ingredients in our SMBH formation and evolution models are the followings: a) the formation of the SMBH particle and in particular the choice to the initial seed mass \citep[e.g.][]{Begelman2006,Volonteri2010}, b) the dynamics of the SMBH particle, with the possible inclusion of a drag force \citep[see the recent work of][]{Tremmel2015}, c) the growth of the SMBH particle mass as a function of time, with two fundamental ingredients being the Bondi-Hoyle-Lyttleton \citep{Hoyle1939, Bondi1944, Bondi1952} accretion rate, limited to the Eddington accretion rate \citep[for observational constrains see e.g.][]{Kollmeier2006, Steinhardt2010}, and finally, d) the feedback from the SMBH particle that affects the surrounding gas \citep{Ostriker2010, Angles-Alcazar2013, Choi2014, Choi2014b, Nayakshin2014, Costa2014}, and therefore couples back to all the previous ingredients of the model. 

In this work, we present a new implementation of the SMBH formation and evolution model in the \ramses{} AMR code \citep{Teyssier2002}, inherited from the earlier work of \cite{Dubois2010} and \cite{Teyssier2011}, but significantly improved in many aspects (see Section~\ref{sec:models}). For example, our sink particle formation sites are automatically extracted from the simulation using the recently developed clump finder onboard the \ramses{} code \citep{Bleuler2014}. We also improved the dynamical integrator of the sink particle, allowing us to perform detailed dynamical studies. Finally, we added two new ingredients to the model, namely a fully momentum conserving drag force and a model for SMBH and Nuclear Star Cluster (NSC) co-evolution. Our goal is to apply these various ingredients to model simulations featuring a cooling, Milky Way-sized halo (See Section~\ref{sec:setup}), leading to the formation of a gas-rich, clumpy and violently turbulent disc, reminiscent of the high-redshift galaxies population detected in deep Hubble Space Telescope images. In Section~\ref{sec:results}, we outline  the fact that SMBH dynamics in this turbulent environment is extremely chaotic, leading to the ejection of the SMBH from the central region of the galaxy, unless one considers very specific dynamical models. Realistic stellar and AGN feedback models make the situation even more critical. In Section~\ref{sec:discussion}, we finally discuss a model where SMBHs are either hosted and protected by a parent NSC, or massive enough to sustain the violent perturbations from their host galaxy. In Section~\ref{sec:summary}, we discuss various observational arguments in favour of this new scenario.
	

\section{A new model for SMBH formation and evolution}\label{sec:models}
 
The first generation of SMBH models was developed in the context of cosmological simulations, with resolution around $ 1~\rm{kpc}$ or more \citep{Bellovary2010, Vogelsberger2014, Schaye2015, Dubois2016} or for relatively smooth galaxy models, using either a pressurised ISM equation of state \citep{Truelove1997, vandeVoort2011} or a low gas fraction relevant for low-redshift galaxy evolution. 
The sink particle was not allowed to move away from the galaxy centre, by either forcing it to remain close to the gravitational potential minimum, or by using various drag forces \citep{Springel2005, Okamoto2008, Gabor2013}.
The next generation of SMBH models need to be able to resolve the SMBH dynamics within the galaxy, and more importantly, to follow its evolution within highly turbulent, gas-rich environments typical of galaxy evolution at high redshift. In this section, we present the new-generation SMBH model implemented in the \ramses{} code. It is heavily based on the old model presented in \cite{Dubois2010} and \cite{Teyssier2011}, and capitalises over the new sink particle implementation we have developed within the context of star-forming molecular clouds \citep{Bleuler2014}. 
 
Although we model SMBHs as collisionless particles, we do not use the Particle Mesh solver designed for the dark matter component. 
Instead, we place around each sink a spherical uniform distribution of test particles (we call them ``cloud particles") of radius $r\sub{sink}~=~4\Delta x\sub{min}$, where $\Delta x\sub{min}$ is the size of a cell at the highest refinement level. These cloud particles are evenly spaced within the sphere (with roughly 8 cloud particles per grid cell) and follow the sink particle as a rigid body.
These cloud particles are used to probe the gas distribution around the sink and to distribute the accretion and the ejection of mass, momentum and energy. Note that the value for the sink sphere radius
can be modified by the user, with recommended values ranging from $1$ to $4\Delta x\sub{min}$. 

In the following subsections we give more details on the improvements of our SMBH sink particle implementation.

\subsection{SMBH formation}\label{ssec:formation}

The life of the SMBH in our simulations begins with the formation of the sink particle. It is a problem which deserves its own careful consideration, but here we reduce it to the identification of a possible formation site and to the choice of the initial mass $M\sub{seed}$.  The two main scenarios for SMBH formation are 1) direct gas collapse or 2) formation through stellar remnants collisions in a dense stellar system. In both case, SMBH formation is associated to exceptionally dense regions, probably at very high redshift, with properties leading first to the formation of an intermediate mass black hole, which will accrete gas and grow even more into the SMBH regime. 

Modelling these processes is clearly out of the scope of this paper, as it would require much higher resolution and the addition of physical ingredients that are absent from our simulations, or that are not even really understood today. We therefore directly create our first and only SMBH when the first dense clump of gas forms. This allows the sink particle to evolve in a dense environment, mimicking the early phase of SMBH growth. For this, we use our built-in clump finder \textsc{phew} \citep{Bleuler2015} and form the sink particle in the most massive gas clump at a chosen time (see Section~\ref{sec:setup}). It is worth emphasising that in this formation scenario seed SMBH is trapped in nuclear gas clump; if the SN feedback is included, then the initial host clump is quickly destroyed.

The value of the initial seed mass is rather arbitrary. A typical value of $M\sub{seed}=10^5$ \msun{}, is usually adopted in large-scale hydrodynamical simulations \citep[e.g.][]{Booth2009}. Direct collapse scenarios of SMBH formation do predict seed masses of this magnitude \citep[e.g.][]{Begelman2006}. In this paper, we prefer to adopt a more pragmatic approach and consider the seed mass as a free parameter. The Bondi accretion model we describe in the next section is based on the strong assumption that the sink particle gravity field dominates over the gas self-gravity. A minimum seed mass equal to the simulation minimum Jeans mass appears to be the right choice, as our numerical experiments in Section~\ref{sec:results} indicate.

\subsection{SMBH accretion}\label{ssec:evolution}

Once the SMBH has formed, it continues to grow in mass via accretion of gas from its surroundings. {Spatial and temporal scales related to accretion process are far from being resolved in all simulations focusing on galactic environments. This motivates the need for sub-grid modelling of the accretion process.}
The most popular 
approach to compute the accretion rate onto the SMBH particle is to use the Bondi-Hoyle-Lyttleton formulae 
\citep[later Bondi for short;][]{Hoyle1939,Bondi1944, Bondi1952}:
\begin{equation}
\dot{M}\sub{Bondi}=4\pi \rho_\infty r\sub{Bondi}^2v\sub{Bondi},\label{eq:bondi_hoyle}
\end{equation}
where 
\begin{eqnarray}
\rho_\infty=\frac{\bar{\rho}}{\alpha(x\sub{sink})} \label{eq:rho_infty}
\end{eqnarray}
$\alpha$ is the dimensionless density profile of the Bondi self-similar solution \citep[see e.g. Chapter 6 of][]{Shu1992}, $\bar\rho$ is the average gas density within the sink sphere, and 
\begin{equation}
x\sub{sink}= r\sub{sink}/r\sub{Bondi}
\end{equation}
is the dimensionless radius evaluated at the sink sphere radius.
This function $\alpha$, first introduced by \cite{Krumholz2004} in tabulated form, is a crucial ingredient to describe the accretion flow, and is often missing in many sink particle algorithm implementations. 
The Bondi radius $r\sub{Bondi}$ and the Bondi velocity $v\sub{Bondi}$ are defined as follows 
\begin{eqnarray}
r\sub{Bondi}&=&\frac{G M\sub{sink}}{v\sub{Bondi}^2},\\
v\sub{Bondi}&=&\sqrt{c\sub{s}^2+v\sub{rel}^2},\label{eq:v_bondi}
\end{eqnarray}
where $c\sub{s}$ is the local sound speed of the gas and $v\sub{rel}$ is the relative velocity between the sink velocity 
$v\sub{sink}$ and the gas average velocity within the sink sphere $\bar v$
\begin{equation}
{\bf v}\sub{rel} = {\bf v}\sub{sink} - \bar{\bf v}
\end{equation}
One can define the free-fall velocity onto the sink particle as
\begin{equation}
v\sub{ff,sink}=\sqrt{\frac{GM\sub{sink}}{r\sub{sink}}}
\end{equation}
The dimensionless radius can be written as
\begin{equation}
x\sub{sink}= v\sub{Bondi}^2/v\sub{ff,sink}^2\label{eq:xsink}
\end{equation}
and obviously indicates whether the accretion flow around the sink is supersonic for $x\sub{sink} < 1$ or subsonic for $x\sub{sink} > 1$.  

In the strong supersonic regime where  $x\sub{sink} \ll 1$, the dimensionless density profile of the Bondi solution asymptotes to
$\alpha(x) \simeq x^{-3/2}$ {(without any underlying assumptions for the equation of state of gas)}. One can re-write the accretion rate in the strong supersonic limit as
\begin{equation}
\dot{M}\sub{Bondi}\simeq4\pi \bar\rho r\sub{sink}^{3/2}\sqrt{G M\sub{sink}} = 3 \frac{M\sub{gas}}{t\sub{ff,sink}}\,
\end{equation}
where the sink free-fall time is defined as 
\begin{equation}
t\sub{ff,sink} = \sqrt{\frac{r\sub{sink}^3}{GM\sub{sink}}} = \frac{r\sub{sink}}{v\sub{ff,sink}}\,
\end{equation}
and the available gas mass within the sink sphere radius is
\begin{equation}
M\sub{gas}=\frac{4\pi}{3}\bar\rho r\sub{sink}^3
\end{equation}
One concludes that in the strong supersonic limit, the accretion rate does not depends on the gas properties anymore,
but only on the available gas mass and the sink free-fall time. This corresponds to a maximum physically motivated accretion rate onto the sink. 

In the strong subsonic limit, where $x\sub{sink} \gg 1$, one has $\alpha(x) \simeq 1$, and the accretion rate can be written as
\begin{equation}
\dot{M}\sub{Bondi} \simeq 4\pi \bar\rho r\sub{Bondi}^{2} v\sub{Bondi}
\end{equation}
This is this formula that is used in most sink particle implementation, and we would like to stress, as in \cite{Krumholz2004},
that this last formulae is only valid in the subsonic regime, where the Bondi radius is much smaller than the sink radius.
Manipulating slightly the previous equation, one can rewrite the accretion rate formulae as
\begin{equation}
\dot{M}\sub{Bondi} \simeq 3 \frac{M\sub{gas}}{t\sub{ff,sink}}\frac{1}{x\sub{sink}^{3/2}}
\end{equation}
This shows explicitly that the subsonic accretion rate is much smaller than the supersonic one. {The transition between the two regimes will of course depend on $v\sub{Bondi}$, $M\sub{sink}$ and the 
adopted resolution ($R\sub{sink}$). Assuming for example that $v\sub{sink}=c\sub{s}=10\unit{km/s}$ and $R\sub{sink}=100\unit{pc}$, then accretion will become supersonic if $M\sub{sink}\gtrsim 4.6\times10^6$~\msun{} and increases to $M\sub{sink}\gtrsim 2.3\times10^8$~\msun{} for $c\sub{s}=100\unit{km/s}$. We would like to emphasise that in the cold accretion regime $R\sub{Bondi}$ is always resolved provided that seed mass is chosen accordingly to the resolution (see \autoref{ssec:results_smbh}).}

{Then it can be also seen} that if the simulation time step is controlled by the sink Courant condition
\begin{equation}
\Delta t \le \frac{t\sub{ff,sink}}{3}\,
\end{equation}
one cannot remove more than the available gas mass within the sink sphere in one time step.

It has been proposed by \cite{Springel2005} and \cite{Booth2009} to boost the previous accretion rate formula, to account for unresolved density and temperature
fluctuations at scales lower than the cell size. In this paper, we follow the same idea, allowing the sound speed of the gas to be reduced, owing to smaller unresolved temperature
fluctuations. This boils down to replacing in the previous formulae the sound speed by
\begin{equation}
c\sub{s} \rightarrow c\sub{s} / \beta\sub{boost}(\bar\rho)
\end{equation}
where the boost factor is defined as
\begin{equation}
\beta\sub{boost}(\rho) = \mathrm{max}[(\rho/\rho_{*})^{2/3},1.0],\label{eq:boost},
\end{equation}
{where $\rho_{*}$ is the critical gas density for star formation (see \autoref{eq:press_floor} in the next section). }

In case of zero relative velocity, this formula corresponds exactly to the model proposed by Booth and Schaye. We would like to stress that the only effect of this boost 
is to change the transition from supersonic to subsonic accretion, but the strong supersonic accretion rate will not be modified from its maximally physically allowed value derived above.
We would like also to stress that one cannot modified the relative velocity $v\sub{rel}$ from physical grounds. Sink particles with very high relative velocities are therefore likely to accrete 
very little mass, as they should. Reducing the relative velocity artificially has been also used in the past to boost the accretion rate, without any physical motivation.

An important ingredient specific to SMBH accretion is the maximal allowed accretion rate onto the black hole, namely the Eddington rate,
\begin{equation}
\dot{M}\sub{Edd} = \frac{4\pi GM\sub{sink}m\sub{p}}{\epsilon\sub{r}\sigma\sub{T} c} = \frac{M\sub{sink}}{t\sub{S}}\label{eq:edd}
\end{equation}
where $m\sub{p}$ is the proton mass, $\sigma\sub{T}$ is the Thomson cross section and $\epsilon\sub{r}$ is the \citet{Shakura1973} radiative efficiency for a SMBH accretion; $\epsilon\sub{r}=0.1$. 
These constants are combined into the Salpeter time, as $t\sub{S} \simeq 45~{\rm Myr}$. 
Finally, the accretion rate onto the SMBH is computed using
\begin{equation} 
\dot{M}\sub{acc}=\mathrm{min}(\dot{M}\sub{Bondi},\dot{M}\sub{Edd}).
\end{equation} 
We would like to stress that the Eddington rate comes from the following picture: gas is accreted using the Bondi rate towards the SMBH accretion disc, 
and the accretion energy is converted into accretion luminosity, which in turn will remove the fully ionised gas in the vicinity of the SMBH, if it exceeds the Eddington luminosity.
Since our accretion model is applied to very large {scales of galactic ISM} (say between 10~pc to 1000~pc), we do not resolve the region where radiation pressure will remove the gas and 
control the accretion onto the SMBH. Eddington limited accretion therefore means that gas is accreted at the Bondi rate, and then decreted at a slightly smaller rate,
the net budget being the (small) Eddington rate. This picture is quite different from what is considered usually and will be used later in the paper to introduce an additional gas drag force on the sink particle.

We discuss finally one important technical detail: once we know the sink particle's current accretion rate, we remove gas from the sink sphere 
by integrating the previous accretion rate over the time step.
\begin{equation} 
\Delta M\sub{gas} = - \dot{M}\sub{acc} \Delta t
\end{equation} 
In order to avoid emptying very low density gas cells in the sink sphere, we remove from each cell (labelled $i$) the following mass-weighted contribution,
\begin{equation} 
\Delta \rho_{i} = - \rho_{i} \frac{\Delta M\sub{acc}}{M\sub{gas}}
\end{equation}
An important consequence of this strategy is that the centre of mass of the accreted gas within the sink sphere does not coincide with the centre of the sphere ${\bf x}\sub{sink}$.

{The Bondi accretion model adopted here is very popular, in both cosmological simulations and star formation communities, because of its great simplicity, which is a strength and a weakness.
It completely ignores the role of angular momentum, turbulence and additional physical effects such as the multiphase and magnetised nature of the ISM in the SMBH vicinity.
Although one can argue that these effects reduce the actual accretion rate on the SMBH, \cite{Negri2016} have shown that Bondi accretion can both lead to over- and underestimating of the SMBH growth, depending e.g. on resolution. Moreover, \cite{RosasGuevara2015} showed recently that modifications to the Bondi formulae implementing the effect of angular momentum 
have no influence in galaxies larger than $10^{11.5}$~\msun{}, like the one we study here (see \autoref{sec:setup}). 
In the present work, we are aiming at growing SMBHs as rapidly as possible, in order to help the sink particle remain on stable central orbits, 
so that the Bondi formula would provide us with an optimistic model,
especially when the sink still has a low mass and resides inside cold and dense gas clumps. 
When the sink mass is larger, and the gas around it becomes hot and diffuse, the role of turbulence, non-radial motion and magnetic fields becomes less important, 
so that the Bondi approach recovers its general validity. }
	
\subsection{SMBH dynamics}\label{ssec:motion}

The next fundamental requirement of our sink particle algorithm is to model properly the dynamics of the SMBH.
The sink particle trajectory follows from the dynamical evolution of a point mass particle, subject to the gravitational force of the gas, stars and dark matter particles, and also subject to a drag force
due to a tight coupling between the accreted gas and the sink. 

Note that the latter has been often invoked in the literature to justify why one could artificially locked the sink particle coordinates to the minimum of the potential well \citep[e.g.][]{Sijacki2007,Costa2014}, or artificially pushed in the direction of the halo centre \citep{Gabor2013}. There is no physical motivation for these models. Lower mass SMBH can be expected to get scattered by massive gas clumps \citep[e.g.][]{Gabor2013}. Other physically motivated models do exist in the literature, that can help preventing the sink particle from wandering around the galaxy. For example,
\citet{Tremmel2015} proposed to estimate the amount of dynamical friction that is missing due to poor resolution, which consists in a sub-grid model for a drag force between the sink and the collisionless component. Similar sub-grid model can be constructed for the potentially missing drag force between the sink and the surrounding gas medium \citep{Chandrasekhar1943, Ostriker1999, Chapon2013}. We will propose here another physically motivated model based on the Eddington limited accretion. 

First, the gravitational interaction between the sink and the matter distribution, as well as between the sink and possible other sinks in the computational box, are both treated
using a direct summation method of a softened $1/r^2$ Newtonian acceleration. We prefer this new approach than using the Particle Mesh method, as it gives more accurate trajectories, especially if the SMBH mass dominates the local potential. The softening radius used in the force calculations is set to $2\Delta x\sub{min}$, as in \citet{Bleuler2014}.
	
When the sink accretes gas from within the sink sphere, it also accretes the corresponding momentum, which translates into an effective drag force between the gas and the sink.
When the accretion rate onto the SMBH is Eddington-limited, the situation is however more complicated. As described before, the Eddington limit for the radiation is enforced in the vicinity of the SMBH, where
the gas is fully ionised and has reached the SMBH accretion disc.  We consider in this paper that the gas accretion rate towards the SMBH accretion disc is set by the Bondi formula, and corresponds to the large scale flow, while the gas accretion rate onto the SMBH is set by the Eddington limit. The difference between the two rates, Bondi minus Eddington,  corresponds to gas being decreted from the accretion disc region and redistributed on large scale, in our case within the sink sphere.
\begin{equation}
\dot{M}\sub{dec}=\dot{M}\sub{Bondi} - \dot{M}\sub{acc}
\end{equation} 
This process of accretion and ejection will lead to an additional exchange of momentum between the gas and the sink, hence an additional drag force. 

We model this additional drag force by requiring that the centre of mass of the joint gas + sink system remain fixed during the accretion, and that its total momentum is conserved. 
If we note the gas centre of mass within the sink sphere as $\bf{x}\sub{gas}$, this translates into a shift in the sink coordinates given by
\begin{eqnarray}
M\sub{gas} \frac{\upright{d}\vect{x\sub{gas}}}{\upright{d} t} & = & \dot{M}\sub{dec} \vect{x\sub{sink}} - \dot{M}\sub{Bondi} \vect{x\sub{gas}}, \nonumber\\
M\sub{sink} \frac{\upright{d}\vect{x\sub{sink}}}{\upright{d} t} & = & \dot{M}\sub{Bondi} \vect{x\sub{gas}} - \dot{M}\sub{dec} \vect{x\sub{sink}},\label{eq:drag1}
\end{eqnarray}
and a similar momentum transfer between the sink and the gas (in other words a drag force) given by:
\begin{eqnarray}
M\sub{gas} \frac{\upright{d}\vect{v\sub{gas}}}{\upright{d} t} & = & \dot{M}\sub{dec} \vect{v\sub{sink}} - \dot{M}\sub{Bondi} \vect{v\sub{gas}}, \nonumber\\
M\sub{sink} \frac{\upright{d}\vect{v\sub{sink}}}{\upright{d} t} & = & \dot{M}\sub{Bondi} \vect{v\sub{gas}} - \dot{M}\sub{dec} \vect{v\sub{sink}},\label{eq:drag2}
\end{eqnarray}
These equations are solved for each time step, and are used to modify the sink position and velocity,
but also the gas density, momentum and total energy within the sink sphere. 
More details on the numerical implementation are given in the Appendix.
Note that in case of zero decreted mass (pure unlimited Bondi accretion), the momentum transfer only comes from the accreted gas mass onto the sink, as it should.
In the opposite case, when the accretion rate is strongly Eddington limited, the mass decretion rate is maximal and almost equal to the Bondi rate. This results in a strong drag force
between the sink and the gas.
	
\subsection{SMBH feedback}\label{ssec:feedback}
	
In this paper, we only consider a model for which thermal energy is injected within the sink sphere, using for the SMBH luminosity the following formula
\begin{equation}
L\sub{AGN} = \epsilon\sub{c} \dot{M}\sub{acc} \epsilon\sub{r}c^2,
\label{eq:e_blast}
\end{equation}
where $\epsilon\sub{r}=0.1$ is the accretion disc radiative efficiency and $\epsilon\sub{c}$ is a free parameter representing the coupling efficiency between the blast wave energy at small scale 
and the resulting thermal energy deposited at large scale. Based on previous work using the \ramses{} code \citep{Teyssier2011, Dubois2012}, we fixed its value to $\epsilon\sub{c} = 0.15$, 
which is quite typical of the corresponding literature, with values ranging  from $0.05$ \citep{Springel2005, Wurster2013} to $0.15$ \citep{Booth2009, Gabor2013}. 

An important improvement compared to the previous \ramses{} implementation is that we deposit now thermal energy 
at every \emph{fine} timestep (i.e. the timestep of the maximum level of refinement $\ell\sub{max}$), and not only at main coarse time steps as before.
We also do not consider a minimum injection temperature, as in \cite{Booth2009} or \cite{Teyssier2011}. 
Moreover, the thermal energy is distributed in every gas cell within the sink sphere proportionally to the gas density. This mass-weighted deposition scheme prevents
the apparition of unrealistically large gas temperature, as opposed to the volume-weighted deposition scheme.

These important changes now allow us to model the competition between heating and cooling within the sink sphere. 
Indeed, one can write an energy equation for the average gas specific internal energy within the sink sphere as
\begin{equation}
\rho \frac{\upright{d} \epsilon}{\upright{d} t} = \frac{L\sub{AGN}}{V\sub{sink}} - n^2 \Lambda(T)
\label{energy_eq}
\end{equation}
where the specific internal energy is related to the temperature and the sound speed by
\begin{equation}
\epsilon \simeq \frac{k\sub{B}T}{\mu m\sub{H}} \simeq c_s^2(t)
\end{equation}
{and $V\sub{sink}$ is the volume of sink accretion zone, $n$ is the gas density in units of H/cc, and $\Lambda$ is a temperature-dependent cooling rate per number density.}

We want now to distinguish two regimes of accretion on the sink. First, we have the cold accretion regime, 
for which cooling dominates over heating. The Bondi accretion rate is so high that we consider the accretion to be Eddington limited,
\begin{equation}
\dot M\sub{acc} = \frac{M\sub{sink}}{t_{\rm S}}.
\end{equation}
We consider for the cooling function only Bremsstrahlung so that
\begin{equation}
\Lambda(T) = \Lambda\sub{0}T^{1/2}
\end{equation}
where $\Lambda_0 \simeq 1.2 \times 10^{-27}$ erg~s$^{-1}$~cm$^3$~K$^{-0.5}$.
This is a good approximation for high temperature and low metallicity gas.
We conclude immediately that, for a given average gas density within the sink sphere, cooling will always win over heating, and the sink will remain in the cold accretion regime,
unless the SMBH mass becomes large enough, so that 
\begin{equation}
M\sub{sink} > n\sub{H}^2 \Lambda_0 T^{1/2} \frac{t\sub{S}}{\epsilon\sub{c}\epsilon\sub{r}c^2} V\sub{sink}.
\end{equation}
Because the sink is now massive enough, heating dominates over cooling, and the sink sphere enters the second phase, 
namely the hot accretion regime. For this, we now assume that the gas temperature is always large enough that the accretion rate is equal to the Bondi rate.
We also consider the SMBH to be at rest in the centre of the galaxy. 
We then obtain for the accretion rate
\begin{equation}
\dot M\sub{acc} \simeq 4\pi \rho \frac{\left( G M\sub{sink}\right)^2 }{c_s^3(t)}
\end{equation}
We can now solve the energy equation, ignoring the cooling term, and obtain the time evolution of the sound speed within the sink sphere
\begin{equation}
c_s(t) = \left[ \frac{15}{2}\epsilon_c \epsilon_r c^2  \left( \frac{G M\sub{sink}}{r\sub{sink}}\right)^2 \frac{t}{r\sub{sink}}\right]^{1/5}
\label{eq:cstime}
\end{equation}
Obviously, the temperature in the sink region will not grow indefinitely. 
As soon as it reaches a high enough value, the gas in the vicinity of the SMBH will expand and cool adiabatically.  
We consider that we have reached the maximum temperature after one sound crossing time of the sink sphere, namely
$t\sub{cross}(t) = r\sub{sink} / c_s(t) = t$.
Combining this with the previous equation gives us the maximum possible sound speed in the hot accretion phase
\begin{equation}
c_{s,\rm max} = \left[ \frac{15}{2}\epsilon_c \epsilon_r c^2  \left( \frac{G M\sub{sink}}{r\sub{sink}}\right)^2 \right]^{1/6}
\label{eq:csad1}
\end{equation}
It can be compared to the galaxy escape velocity to assess the possibility for the SMBH to unbind the gas from the nuclear region (see below).

Besides various constants that we set to our fiducial values ($\epsilon_c=0.15$ and $\epsilon_r=0.1$), we see that the only variables entering theses various formulae are the SMBH mass, 
$M\sub{sink}$, the sink sphere radius $r\sub{sink}$, and finally the average gas density within the sink sphere $n_{\rm H}$.
Inserting typical values for our present simulation, we can compute first the critical SMBH mass beyond which heating dominates over cooling,
so that the sink sphere can exit the cold accretion regime and actually heats the gas around the SMBH
\begin{equation}
M\sub{sink,crit}^{\rm cool} \simeq 8 \times 10^4~M\sub{\odot}~\left( \frac{n_{\rm H}}{100~{\rm H/cc}} \right)^2 \left( \frac{r\sub{sink}}{100~{\rm pc}} \right)^3
\label{eq:msinkcrit_cool},
\end{equation}
where we assumed the gas temperature to be fixed at $10^6$~K in the cooling function. 
If this is the case, then the temperature within the sink sphere will steadily increase according to \autoref{eq:csad1} and reach the maximum sound speed
\begin{equation}
c\sub{s, max} \simeq 750~{\rm km/s}~\left( \frac{M\sub{sink}}{10^8~M\sub{\odot}}\right)^{1/3} \left( \frac{r\sub{sink}}{100~{\rm pc}}\right)^{-1/3}\label{eq:csad}
\end{equation}
{This last equation can be used to define another critical mass, $M\sub{sink,crit}^{\mathrm{esc}}$, corresponding to $c\sub{s, max}=v\sub{esc}$, the escape velocity from the centre of the halo,
so that AGN heating would result in the unbinding of the hot gas in the vicinity of the SMBH. We find 
\begin{equation}
M\sub{sink,crit}^{\mathrm{esc}} = 10^8\,\mathrm{M}_\odot \left( \frac{v\sub{esc}}{750~{\rm km/s}} \right)^3 \left( \frac{r\sub{sink}}{100~{\rm pc}} \right)
\label{eq:msinkcrit_vesc}
\end{equation}

}
In summary, if enough gas makes it into the sink sphere, the density will be high and 
cooling will dominate,  maintaining the gas temperature to relatively low values and the accretion rate to the Eddington limit. 
If, on the other hand, the gas density within the sink sphere is too low, or if the sink mass is too large, we enter the hot, adiabatic regime 
for which the gas temperature is quickly rising to its maximum value. 
Unfortunately, as we will see in the Results section, all these quantities depend sensitively on the adopted resolution. 
A better spatial resolution, resulting in a smaller sink radius,
can reduce the critical SMBH mass, but can also increase it by allowing for larger gas densities. 
Better spatial resolution can also increase the gas temperature in the hot accretion regime significantly.

{On the other hand, we could also apply the same formalism to the ISM in the vicinity of the SMBH, using the fundamental properties of a realistic multiphase gas rather than the 
relatively artificial properties of our finite resolution simulations.
For example, one can relate the gas density in the cooling critical mass formula to the average density of typical gas clouds that are bombarding the SMBH in the nuclear region, and one can argue that the feedback energy should be deposited within a fixed radius, invoking other physical processes to set this energy deposition scale. 
In what follows, we will only apply our simple analytical arguments to interpret our numerical results, and defer a more general and realistic description of the ISM around the SMBH to future work.}

\section{Numerical setup}\label{sec:setup}
	
We use the AMR code \ramses{} \citep{Teyssier2002} and its second-order, unsplit Godunov scheme to solve the Euler equations. 
The evolution of dark matter and stars is performed with the Adaptive Particle-Mesh solver with cloud-in-cell interpolation. 
The dynamical evolution of the sink particle is performed the direct gravitational acceleration (see \autoref{ssec:motion}).

Our initial conditions feature an isolated, gas-rich, slowly rotating (spin parameter of $0.04$) dark matter halo of $2\times10^{12}$ \msun{} sampled using one million dark matter particles. The halo has a truncated NFW \citep{Navarro1997} profile with a concentration parameter $c=10$ and with the circular velocity $V\sub{200}=160\,\mathrm{km\,s}^{-1}$, which results in the radius $R\sub{200}=230\,\mathrm{kpc}$, while the halo is truncated at $514\,\mathrm{kpc}$.
Initially, the gaseous halo is in hydrostatic equilibrium and has the universal gas fraction of $f\sub{gas}=15\%$. The initialisation follows the setup of \citet{Teyssier2013}. Our fiducial run has a spatial resolution of $\Delta x\sub{min}=78\,\mathrm{pc}$.
	
Using an isolated cooling halo is dictated by a compromise between realistic but expensive cosmological simulations and idealised but highly resolved isolated disc simulations.
Since we are using a realistic initial angular momentum profile inspired from the average angular momentum distribution from N body simulation \citep{Bullock2001},
gas will be continuously accreted from the halo into the disc, with the right amount of angular momentum, giving us the possibility to feed the nuclear region, and possibly the central SMBH.
	
We use the \citet{Sutherland1993} model for radiative cooling of gas for H, He and metal lines for gas hotter than $10^4\,\mathrm{K}$ and from metal fine-structure cooling processes at lower temperatures. 
We advect the metallicity in the form of a passive scalar and we choose the initial metallicity to be $Z\sub{ini}=0.05$ Z$\sub{\odot}$.
A pressure floor is introduced at high density and low temperature, to prevent the uncontrolled fragmentation of gas beyond the spatial resolution, possibly leading to the formation of numerical singularities (especially because we are using a low star formation efficiency).
The temperature corresponding to the pressure floor is set to 
\begin{equation}
T\sub{floor}=T\sub{*}\left(\frac{n\sub{\mathrm{H}}}{n\sub{*}}\right)^{\Gamma-1}\label{eq:press_floor}
\end{equation}
with a critical gas number density $n\sub{*}=9\,\mathrm{cm}^{-3}$, a critical temperature $T\sub{*}=2\times10^3\,\mathrm{K}$, and $\Gamma=2$. 
This results in the minimum Jeans length 
\begin{equation}
\lambda\sub{J}=c\sub{s} \sqrt{\frac{\pi}{G\rho}} = \sqrt{\frac{\Gamma \pi k\sub{B}T_*}{m_{\rm H}^2 G n_*}} \simeq 332\, {\rm pc} \simeq 4\,\Delta x\sub{min}
\end{equation}
and in the minimum Jeans mass
\begin{equation}
M\sub{J}=\frac{4\pi}{3} n_* m_{\rm H} \left( \frac{\lambda\sub{J}}{2} \right)^3 \simeq 4 \times 10^6 M\sub{\odot}
\end{equation}
The mesh refinement strategy we have adopted for all our simulations is a quasi-Lagrangian approach, where cells are refined once their mass exceed $8\times m\sub{res}$,
where our mass resolution is set to $m\sub{res} \simeq 1.5 \times 10^5 M\sub{\odot}$, so that our minimum Jeans mass is always sampled by at least 32 resolution elements.
In all simulations, star formation is modelled with a Schmidt law with a rather low efficiency $\epsilon_*=0.01$ coming from observations of local molecular clouds \citep{Krumholz2007}. 
Collisionless star particles of fixed mass $1.3\times10^5$ \msun{} are spawned stochastically with a Poisson distribution if the gas density in the cell is larger than $n\sub{*}m\sub{\mathrm{H}}$ \citep{Rasera2006}.
Feedback from supernovae, if considered, is modelled with a non-thermal energy injection with efficiency of 10\% {(i.e. 10\% of stellar population explodes, each SN with energy of $10^{51}$~erg)} and yield of 10\% (1 M$\sub{\odot}$ of metals for each 10 M$\sub{\odot}$ of ejected material).
The non-thermal energy dissipation timescale is set to 10~Myr. 
We boost the efficiency of our supernovae feedback recipe by grouping stochastically multiple star particles into one single star cluster of mass $10^8$ M$\sub{\odot}$.

As it was already mentioned in Section~\ref{ssec:formation}, we allow only one sink to form in our galaxy. {While star formation and stellar feedback are both modelled since the very beginning, we only form the sink particle at around 200 Myr after the start of the simulation. This time roughly corresponds to the stage in the disc evolution in which massive gas clumps are present and the environment of SMBH is well established in terms of gas and stars. This should promote most stable growth conditions for the newly seeded sink.}
We use the \textsc{phew} clump finder \citep{Bleuler2015} to identify the most massive gas clump {of a mass of order of $10^8$~\msun{}} as the formation site for the SMBH, and let the sink evolve from there. {Initial velocity of the sink corresponds to that of gas out of which it was formed. Mass, momentum and angular momentum are conserved during the formation process.}
All fiducial parameters of our SMBH model are listed in \autoref{tab:parameter_summary}. 
		
\begin{table}
\begin{center}	
\caption{Summary of fiducial parameters related to SMBH sink particles in \ramses{} simulations.}
\label{tab:parameter_summary}
\begin{tabular}{p{0.15\columnwidth}p{0.135\columnwidth}p{0.55\columnwidth}}
\hline
\hline
Parameter & Fiducial value & Description\\
\hline
\hline
$M\sub{seed}$ & $10^6$ M$_\odot$ & Sink seed mass\\\hline
$M\sub{clump}$ & $10^{8}$  M$_\odot$ & Mass of the clump in which we seed the sink\\\hline
Direct solver & yes & The direct $N$-body solver used to evolve the trajectory of a sink\\\hline
Drag & yes & Gas drag force from accretion\\\hline
$\alpha\sub{boost}$ & \autoref{eq:boost} & Boost factor for the Bondi velocity\\
\hline\hline
\end{tabular}
\end{center}
\end{table}

\section{Results}\label{sec:results}

We now present our simulation results, including each important process one by one, in order to compare them, and gauge their relative importance.
These processes are listed in \autoref{tab:parameter_exploration}. 
For each feedback process, we use the parameters described in the previous section. We however consider the SMBH seed mass as a free parameter, and we explore values ranging from $10^5$
to $10^9$~M$_\odot$, as listed in  \autoref{tab:parameter_exploration}. 

\begin{table}
\begin{center}	
\caption{Summary of simulation runs and parameters used in this study. Parameters varied with respect to the fiducial run are highlighted in bold print. Columns: (1) subsection in which the simulations are analysed (with exception for fiducial run); (2) maximum allowed refinement level; (3) fraction of SN energy deposited in the gas; (4) drag force modelled (or inclusion of a nuclear star cluster); (5) initial seed mass in $\log_{10}$ \msun{}; (6) AGN feedback.}
\label{tab:parameter_exploration}
\begin{tabular}{ccccccc}
\hline
\hline
Section  & $l\sub{max}$ & $\epsilon\sub{SN}$ & drag & $m\sub{seed}$ & AGN fbk.\\
(1) & (2) & (3) & (4) & (5) & (6)\\
\hline\hline
\ref{ssec:results_smbh}	& 14 & 0.0 & yes &  \textbf{5} & \textbf{no}\\
\ref{ssec:results_smbh}	& 14 & 0.0 & yes &  6 & \textbf{no}\\
\ref{ssec:results_smbh}	& 14 & 0.0 & yes &  \textbf{7} & \textbf{no}\\
\ref{ssec:results_smbh}	& 14 & 0.0 & yes &  \textbf{8} & \textbf{no}\\
\ref{ssec:results_smbh}	& 14 & 0.0 & yes &  \textbf{9} & \textbf{no}\\
\hline
\ref{ssec:results_agn}	& 14 & 0.0 & yes &  \textbf{5} & yes\\
\ref{ssec:results_agn}	& 14 & 0.0 & yes &  6 & yes\\
\ref{ssec:results_agn}	& 14 & 0.0 & yes &  \textbf{7} & yes\\
\ref{ssec:results_agn}	& 14 & 0.0 & yes &  \textbf{8} & yes\\
\ref{ssec:results_agn}	& 14 & 0.0 & yes &  \textbf{9} & yes\\
\hline
\ref{ssec:results_sn_smbh}	& 14	& \textbf{0.1} & yes &  \textbf{5} & \textbf{no}\\
\ref{ssec:results_sn_smbh}	& 14	& \textbf{0.1} & yes &  6 & \textbf{no}\\
\ref{ssec:results_sn_smbh}	& 14	& \textbf{0.1} & yes &  \textbf{7} & \textbf{no}\\
\ref{ssec:results_sn_smbh}	& 14	& \textbf{0.1} & yes &  \textbf{8} & \textbf{no}\\
\ref{ssec:results_sn_smbh}	& 14	& \textbf{0.1} & yes &  \textbf{9} & \textbf{no}\\
\hline
\ref{ssec:results_sn_agn}& 14 & \textbf{0.1} & yes &  \textbf{5} & yes\\
\ref{ssec:results_sn_agn}& 14 & \textbf{0.1} & yes &  6 & yes\\
\ref{ssec:results_sn_agn}& 14 & \textbf{0.1} & yes &  \textbf{7} & yes\\
\ref{ssec:results_sn_agn}& 14 & \textbf{0.1} & yes &  \textbf{8} & yes\\
\ref{ssec:results_sn_agn}& 14 & \textbf{0.1} & yes &  \textbf{9} & yes\\	
\hline	
\ref{ssec:results_sn_nsc}	& 14 & \textbf{0.1} & \textbf{NSC} &  \textbf{5} & yes\\
\ref{ssec:results_sn_nsc}	& 14 & \textbf{0.1} & \textbf{NSC} &  6 & yes\\
\ref{ssec:results_sn_nsc}	& 14 & \textbf{0.1} & \textbf{NSC} &  \textbf{7} & yes\\
\ref{ssec:results_sn_nsc}	& 14 & \textbf{0.1} & \textbf{NSC} &  \textbf{8} & yes\\	
\ref{ssec:results_sn_nsc}	& 14 & \textbf{0.1} & \textbf{NSC} &  \textbf{9} & yes\\		
\hline
\ref{ssec:res_effects}	& \textbf{15} & 0.0 & yes &  6 & yes\\
\ref{ssec:res_effects}	& \textbf{15} & \textbf{0.1} & yes &  6 & yes\\
\ref{ssec:res_effects}	& \textbf{15} & \textbf{0.1} & \textbf{NSC} &  6 & yes\\
\hline\hline
\end{tabular}
\end{center}
\end{table}
	
\subsection{Accretion-limited growth}\label{ssec:results_smbh}

Our first suite of simulations has been performed without any feedback processes and with only one sink particle seeded in the first, massive enough, nuclear gas clump, growing via Eddington-limited Bondi accretion. Because of the relatively low angular momentum in our cooling halo, mimicking what we expect from cosmological simulations, these simulations without feedback lead to the formation of a gas-rich, clumpy and bulge-dominated galaxy \citep[see also][]{Teyssier2013, Dubois2016} that resembles many observed high-z galaxies, in particular the so-called ``blue nuggets" \citep[see e.g.][]{Damjanov2009}. 

\begin{figure*}
\begin{subfigure}{0.45\textwidth}
\centering
\includegraphics[width=\columnwidth]{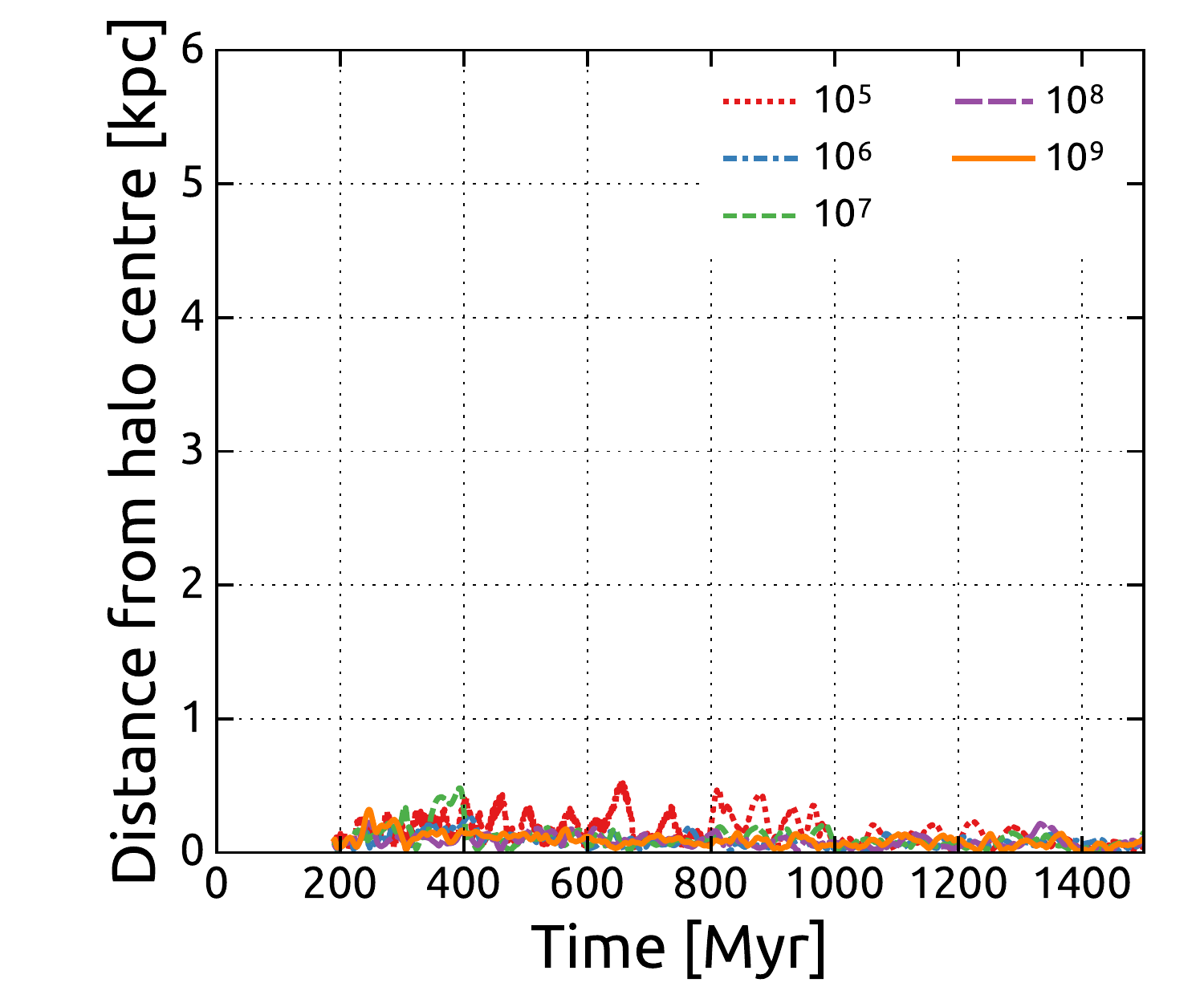}
\caption{Distance between the sink and the centre of the halo}\label{fig:dist_cent_smbh}
\end{subfigure}
\qquad
\begin{subfigure}{0.45\textwidth}
\centering
\includegraphics[width=\columnwidth]{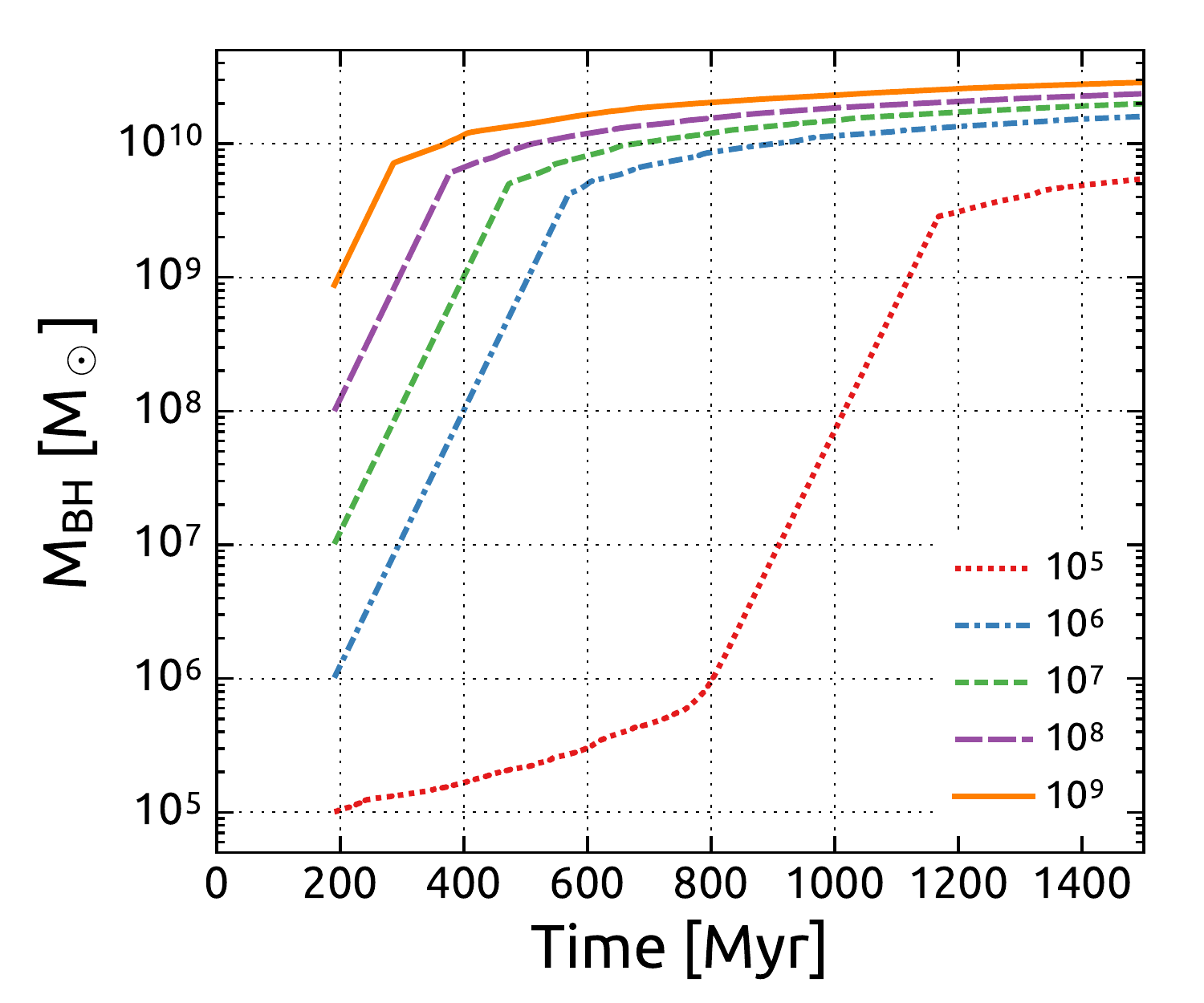}
\subcaption{SMBH mass growth}\label{fig:mass_evo_smbh}
\end{subfigure}
\caption{Evolution of distance to the centre of halo and sink mass for the runs without neither SN and AGN feedbacks for five different seed masses: {$10^5$~\msun{} - red (dotted), $10^6$~\msun{}  - blue (dash-dotted), $10^7$~\msun{}  - green (short dashes), $10^8$~\msun{}  - purple (long dashes), and $10^9$~\msun{}  - orange (solid). The sink particle occupies position in the centre of the halo and its growth is limited first by Eddington rate and later by angular momentum loss in the gas. Lack of AGN feedback heating leads to worrisomely large SMBH mass.}}
\label{fig:hsc_smbh}
\end{figure*}

The trajectory and the mass growth of the SMBH are shown in \autoref{fig:hsc_smbh}. For all our adopted seed masses, the SMBH remains well within the nuclear region (defined here as the central kiloparsec), in which they were born. Interestingly, the lowest seed mass $10^5$~M$_\odot$ shows a very different behaviour than the other, larger seed masses. Its growth is very slow, for almost 1~Gyr,
and only when it reaches $10^6$~M$_\odot$ does it have a high enough accretion rate and grows exponentially. The other seed masses start growing exponentially immediately after their creation, which means that they are massive enough to have a sustained, larger than Eddington, Bondi accretion rate. 

We argue that the critical mass for the sink particle to accrete fast enough is the minimum Jeans mass 
associated to our adopted mesh resolution.
Indeed, assuming that  $v\sub{rel}=0$, we can re-write the parameter that controls whether Bondi accretion is subsonic or supersonic (see Eq.~\ref{eq:xsink}) as
\begin{equation}
x\sub{sink}  =  \frac{c_s^2 \lambda\sub{J}}{GM\sub{sink}}
		   \simeq \frac{M\sub{J}}{M\sub{sink}}
\label{eq:jeans2sink}
\end{equation}
where we used the fact that $r\sub{sink}=4\Delta x\sub{min}=\lambda \sub{J}$.
For our fiducial resolution, the Jeans mass $M\sub{J}$ is $4\times10^6$~M$_\odot$. 
For the lowest seed mass, which is below the Jeans mass, accretion follows the Bondi rate, and is rather low, because the accretion is subsonic. Note that in this regime, because the accretion rate is low, the dynamical coupling between the sink and the gas is weak, making the sink very sensitive to external perturbations. One can see in \autoref{fig:hsc_smbh} that the trajectory of the sink particle is quite perturbed, with visible oscillations around the centre of the galaxy. These oscillations increase the relative velocity between the gas and the sink, further contributing to the low accretion rate.  
Once the sink mass grows beyond $10^6$~M$_\odot$, about 800 Myr after the start of the simulation for the small seed mass or immediately after creation for the other seed masses, 
the Bondi accretion evolves from subsonic to supersonic. 
A much more rapid, Eddington-limited exponential growth follows. 

After this phase, the SMBH mass seems to saturate, and grows only mildly, mostly because of the slow accretion of fresh gas into the nuclear region.
Indeed, since we {did not} include any feedback processes in this first experiment, the final SMBH mass is regulated by the available gas mass within the nuclear region. 
This regime, called here accretion-limited growth, was first discussed in \cite{Bournaud2011}.
The late accretion phase is controlled by angular momentum transfer in the galactic disk, triggered by various instabilities and slowly feeding the SMBH with fresh gas.
In this case, the SMBH trajectory remains well within the nuclear region, a dense and massive stellar bulge that provides a very stable environment for the SMBH.
As a result, the sink particle never leaves the nuclear region.  

\subsection{AGN feedback-limited growth}
\label{ssec:results_agn}

\begin{figure*}
\begin{subfigure}{0.45\textwidth}
\centering
\includegraphics[width=\columnwidth]{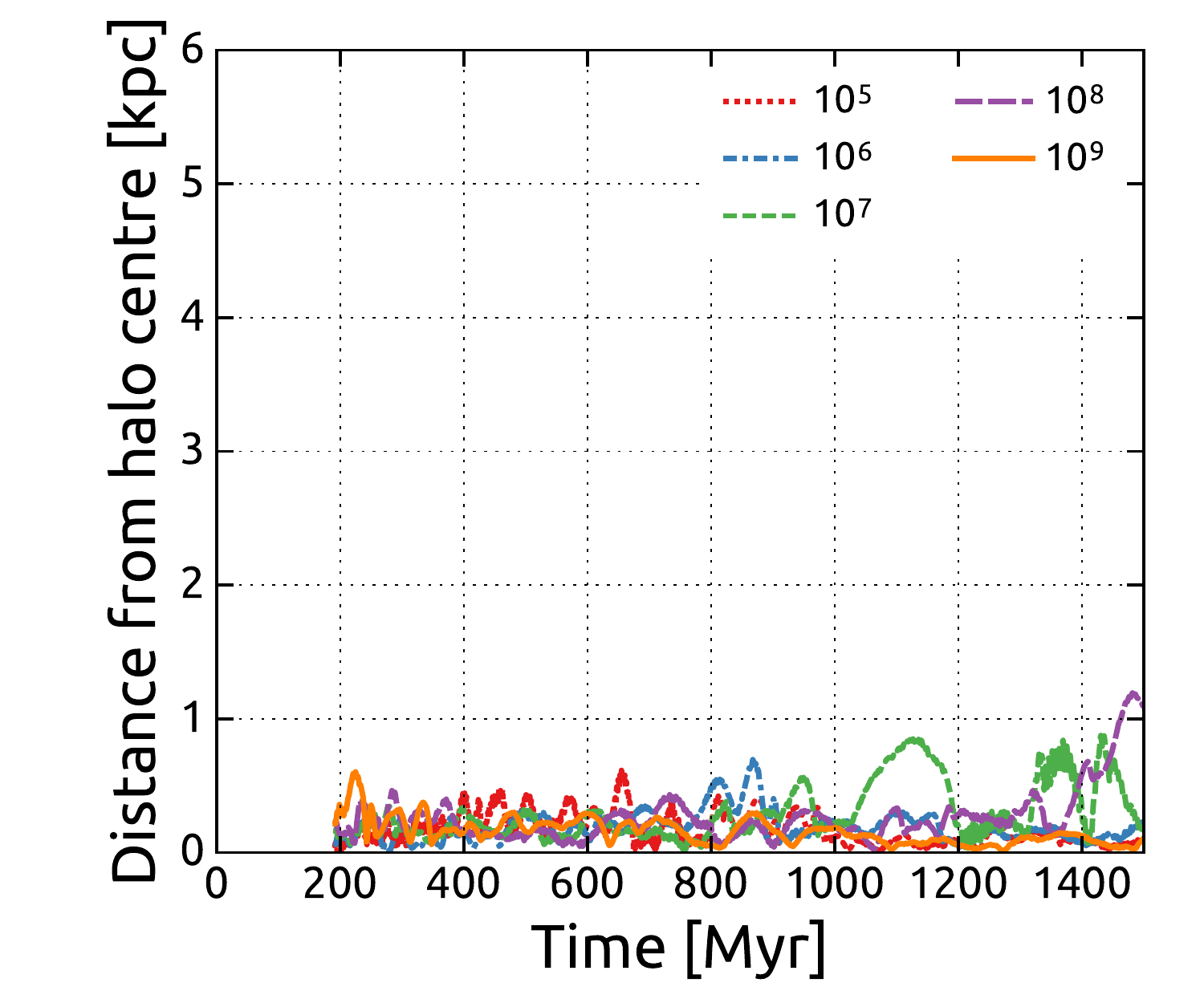}
\caption{Distance between the sink and the centre of the halo}\label{fig:dist_cent_agn}
\end{subfigure}
\qquad
\begin{subfigure}{0.45\textwidth}
\centering
\includegraphics[width=\columnwidth]{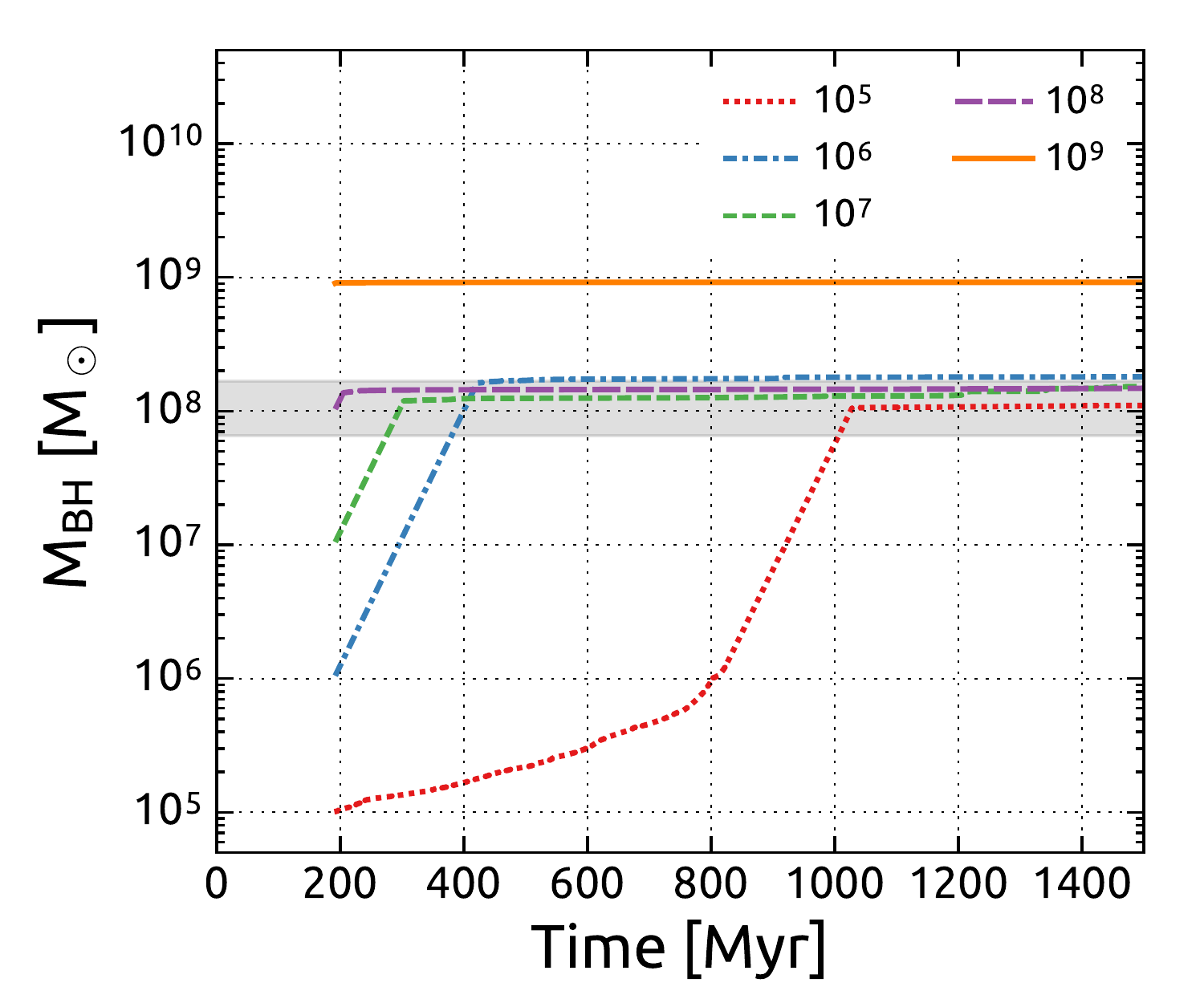}
\subcaption{SMBH mass growth}\label{fig:mass_evo_agn}
\end{subfigure}
\caption{Evolution of distance to the centre of halo and sink mass for the runs with {AGN feedback only for five different seed masses: $10^5$~\msun{} - red (dotted), $10^6$~\msun{}  - blue (dash-dotted), $10^7$~\msun{}  - green (short dashes), $10^8$~\msun{}  - purple (long dashes), and $10^9$~\msun{}  - orange (solid). Grey band on the right panel shows predicted SMBH mass based on the density in the sink sphere (cf. \autoref{eq:msinkcrit_cool}) - lower envelope corresponds to density of 500~H/cc, while upper to 800~H/cc (see \autoref{fig:vesc_agn}). The sink particle resides in the centre of the halo travelling with most massive clump and its growth is limited first by Eddington rate and later terminated at self-regulation scale due to its feedback heating.}}
\label{fig:hsc_agn}
\end{figure*}

We have repeated the same simulations as in the previous section, but this time with AGN feedback. The resulting dynamical and mass evolutions of the SMBH are shown in \autoref{fig:dist_cent_agn} and \autoref{fig:mass_evo_agn}.
The only difference with the previous setup is the final mass of the sink, which is now regulated by AGN feedback. 

The initial growth of the SMBH in our simulations with AGN feedback is very similar to the runs without feedback. Due to the large amounts of gas in the nuclear region, 
feedback heating does no affect the gas surrounding the sink, as cooling dominates. 
As soon as the accretion rate is high enough, heating dominates over cooling and the SMBH quickly reaches its maximum mass, 
which in our case is around $2\times10^8$ \msun{}.

\begin{figure*}
\centering
\includegraphics[width=\textwidth]{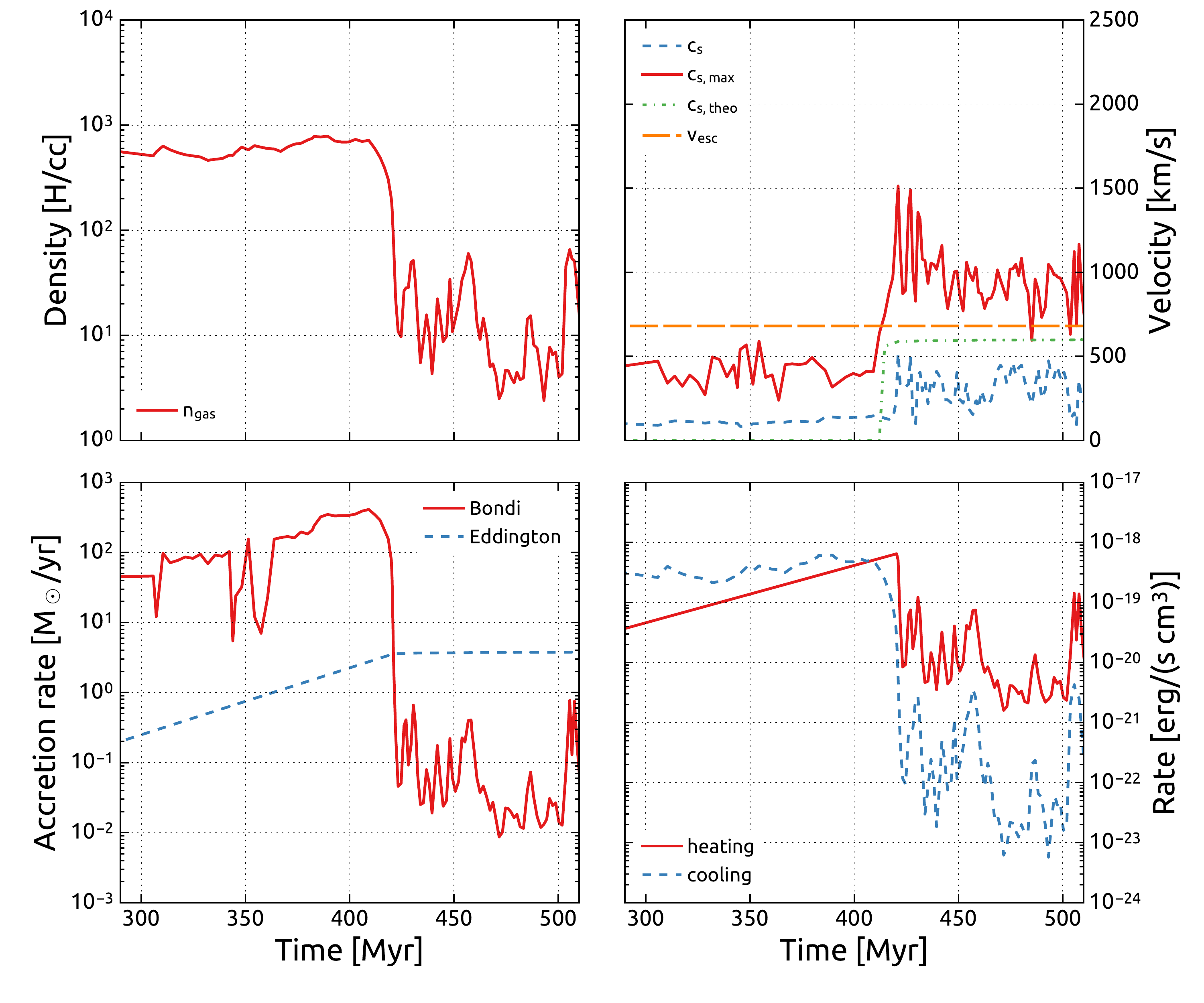}
\caption{Time evolution of 1) average gas density within the sink sphere (\emph{top left}); 2) average, mass-weighted, sound speed {(blue, short dashes) and maximum sound speed (red, solid) within the sink sphere (\emph{top right}), we have also represented our simple theoretical model (Eq.~\ref{eq:csad1} and \ref{eq:csad}) (green, dot-dashed) compared to the escape velocity from the halo's centre (orange, long dashes); 3) Bondi (red, solid) and Eddington (blue, dashed) accretion rates (\emph{bottom left}) and 4)  average heating (red, solid) and cooling (blue, dashed) rates within the sink sphere (\emph{bottom right}) for simulation with AGN feedback only and \mseed{6}.}}
\label{fig:vesc_agn}
\end{figure*}

The maximum, self-regulated mass is related to the heating-cooling balance we have discussed in \autoref{ssec:feedback} (see Eq.~(\ref{eq:csad1}) and (\ref{eq:csad})). 
Using the simulation with \mseed{6} as an example, we see that at around 420~Myr, the sink's growth is terminated. 
Initially, the gas density in the sink sphere is quite large, around $\sim600$~H/cm$^3$, so that clearly cooling dominates the energy budget in the sink sphere.
Gradually, as the SMBH mass grows, feedback is able to heat the gas more and more in the sink sphere, until the SMBH mass reaches the critical value for which heating dominates; {this can be compared with estimate given by \autoref{eq:msinkcrit_cool}, which is plotted as a grey band on \autoref{fig:mass_evo_agn}}.
Very quickly the gas sound speed within the sink sphere rises, until it exceeds the escape velocity of the halo.
When this happens, gas is removed from the nuclear region by a blast wave, which reduces the average gas density down to or even below $\bar{\rho} \simeq 10$~H/cm$^3$, 
and makes feedback even more efficient. 
Feedback is able to maintain the sound speed to a high value ($c\sub{s}\simeq 400$ km/s), strongly reducing the accretion rate (see \autoref{eq:v_bondi} and bottom left panel of \autoref{fig:vesc_agn}).

This is only when feedback processes are able to accelerate gas to the escape velocity that the growth of SMBH is halted \citep[see also][]{Silk1998, Fabian1999}. 
This can be seen on \autoref{fig:vesc_agn}, where we plot various average quantities measured in the sink sphere. 
The gas density (\emph{top left}) drops by two orders of magnitude as soon as the maximum sound speed significantly exceeds the halo escape velocity ($v\sub{esc}$, \emph{top right}).
The critical SMBH mass $M\sub{sink, crit}$, for which heating balances cooling, is reached at 420~Myr (\emph{bottom right}), after which the average sound speed quickly exceeds $v\sub{esc}$, 
which then marks the end of the cold accretion regime (\emph{bottom left}) and the beginning of the hot mode of accretion. For comparison, we have plotted in \autoref{fig:vesc_agn} our simple analytical predictions
from \autoref{ssec:feedback}. We can predict quite nicely the onset of efficient heating, when the SMBH mass reaches its critical value, as well as the end of the mass growth, when the maximum sound speed
reaches the escape velocity of the halo.

The case with \mseed{9} is very different than all the other cases. 
Here, the initial seed mass is already above the maximum, self-regulated mass.
AGN feedback immediately blows away the gas from the nuclear region.
As a result, the gas in the vicinity of the sink remains very hot and the accretion rate very low.
	
\subsection{Supernovae feedback-limited growth}
\label{ssec:results_sn_smbh}

\begin{figure*}
\begin{subfigure}{0.45\textwidth}
\centering
\includegraphics[width=\columnwidth]{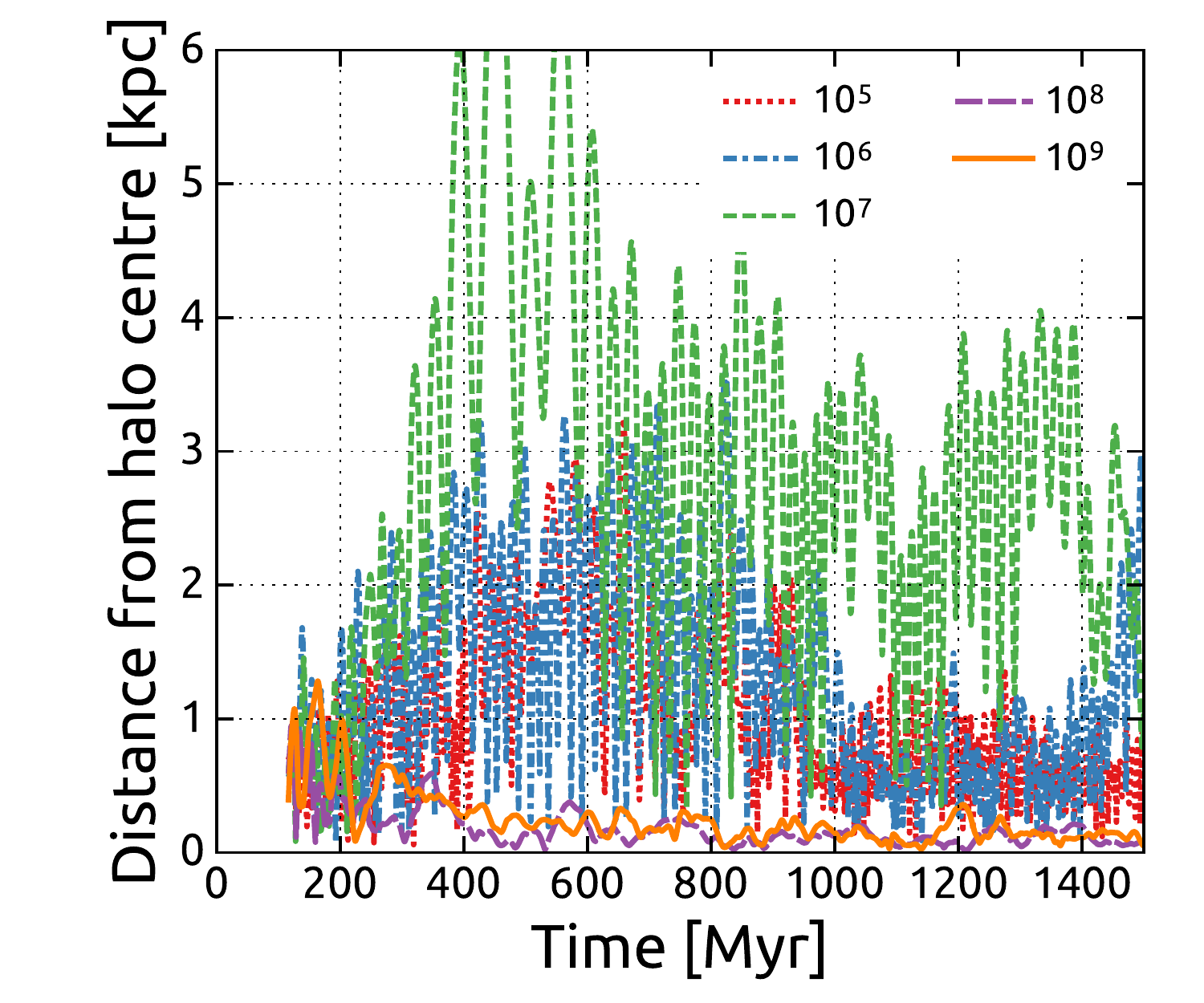}
\caption{Distance between the sink and the centre of the halo}\label{fig:dist_cent_sn_smbh}
\end{subfigure}
\qquad
\begin{subfigure}{0.45\textwidth}
\centering
\includegraphics[width=\columnwidth]{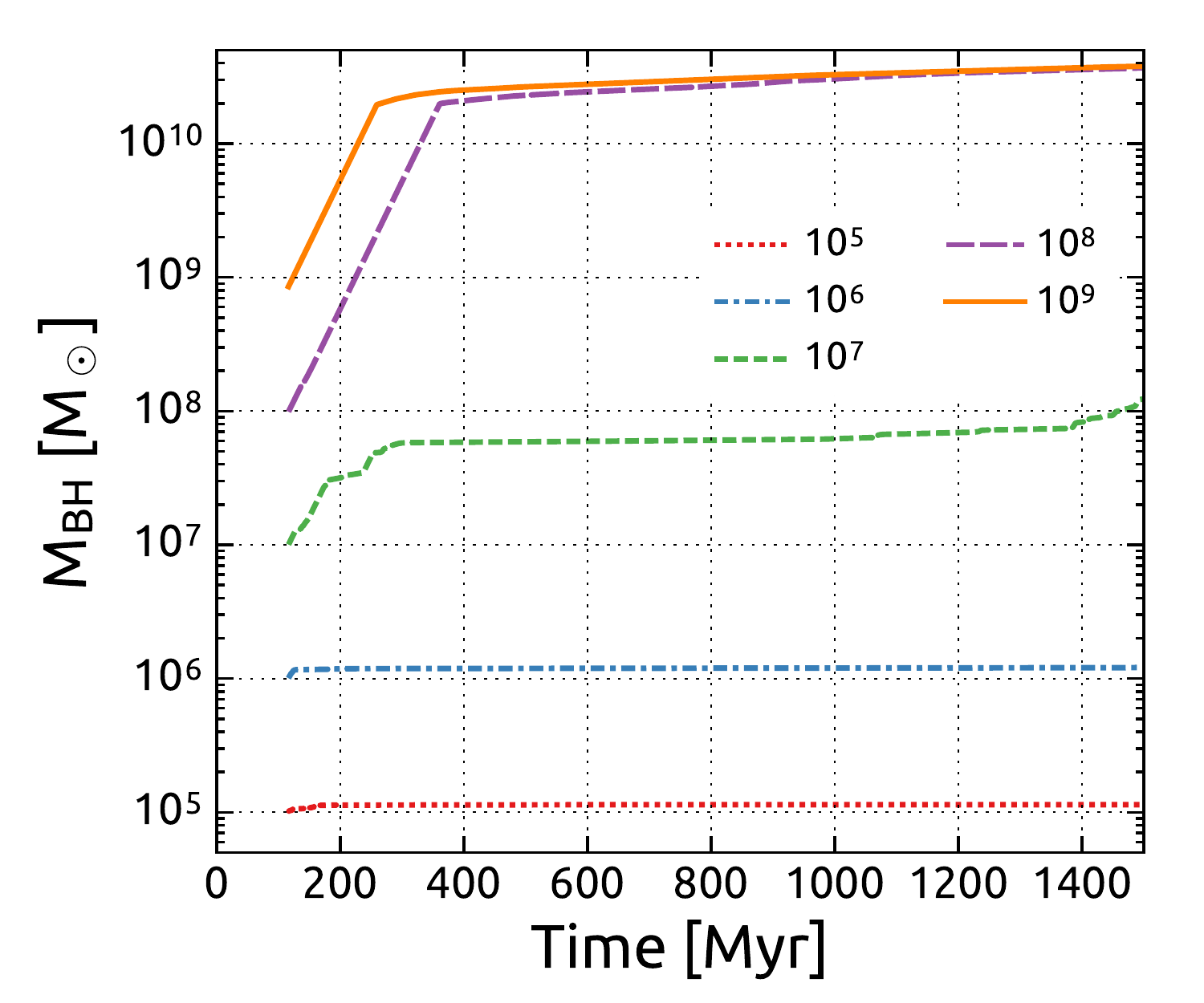}
\subcaption{SMBH mass growth}\label{fig:mass_evo_sn_smbh}
\end{subfigure}
\caption{Evolution of distance to the centre of halo and sink mass for the runs with SN feedback but \emph{without} AGN feedback for five different seed masses: {$10^5$~\msun{} - red (dotted), $10^6$~\msun{}  - blue (dash-dotted), $10^7$~\msun{}  - green (short dashes), $10^8$~\msun{}  - purple (long dashes), and $10^9$~\msun{}  - orange (solid). Grey band on the right panel shows predicted SMBH mass based on the halo escape velocity (cf. \autoref{eq:msinkcrit_vesc}).}}
\label{fig:hsc_sn_smbh}
\end{figure*}

We now remove AGN feedback from the picture, but include instead supernova feedback from dying massive stars.
We use the same simulation suite than before, with seed masses from \mseed{5} to \mseed{9}.
On \autoref{fig:hsc_sn_smbh}, we again plot the time evolution of the distance of the sink particle to the halo centre and of its mass. 
Here again, we can see two different regimes. 
Low and intermediate seed masses are quickly removed from the central kiloparsec. There, supernovae feedback is efficient enough to destroy the parent clump, and the sink particles
are perturbed by interaction with nearby clumps. As a consequence, the trajectory of the sinks becomes more complicated and eccentric, 
and the relative velocity between the sink and the gas within the sink sphere grows significantly, reducing the accretion rate and the corresponding drag force accordingly. 
For seed masses larger than \mseed{8}, the sink trajectory appears as much less perturbed and the sink manage to remain within the nuclear region. As a consequence, accretion proceeds
much more rapidly and the sink mass can grow up to its accretion-limited value, as in Section~\ref{ssec:results_smbh}.

In order to estimate the mass of the typical clumps that will perturb the trajectory of the SMBH, we use the classical Toomre analysis of gas fragmentation
in an idealised razor thin disc \citep{Toomre1964}. The largest unstable wavelength is the Toomre length

\begin{equation}
\lambda\sub{T} 
\simeq 
\frac{G\Sigma\sub{gas}}{\kappa^2} 
\simeq
\frac{f\sub{gas}}{\pi}R\sub{gal}.
\end{equation}
In this approximate formula $G$ is gravitational constant, ${\Sigma\sub{gas}=M\sub{gas}/\pi R\sub{gal}^2}$ is the gas surface density, 
${\kappa \simeq V\sub{gal}/R\sub{gal}}$ is the epicyclic frequency, ${V\sub{gal}=\sqrt{G M\sub{tot}/R\sub{gal}}}$ is the galaxy circular velocity
and $f\sub{gas}$ is the gas-to-total mass fraction in the disc. 
In order for this wavelength to be truly unstable, the Toomre parameter must satisfy
\begin{equation}
Q = \frac{c\sub{s} \kappa}{\pi G \Sigma\sub{gas}} < 1.
\end{equation}
where $c\sub{s}$ can be taken as either the sound speed or the velocity dispersion of the gas. 
Under such conditions, one can then estimate the mass of the most massive clumps as the Toomre mass $M\sub{T}$ defined by
\begin{equation}
M\sub{T}=\Sigma\sub{gas} \pi \left(  \frac{\lambda\sub{T}}{2} \right)^2
\simeq 
\frac{M\sub{tot}f\sub{gas}^3}{4\pi^2}.
\label{eq:mtoomre}\end{equation}
For a Milky Way-like galaxy, one has ${M\sub{tot} \simeq10^{11}}$~\msun{} in the disc (not to be confused with the total mass in the halo, which is one order of magnitude larger).
At low redshift, in galaxies similar to our own Milky Way, one finds ${f\sub{gas} \simeq 0.1}$, which results in a typical clump mass of ${M\sub{T} \simeq2.5\times10^6}$~\msun{}. At high redshift, however,
like the cooling halo set-up we are adopting in this paper, the gas fraction is much higher, ${f\sub{gas} \simeq 0.5}$, for a similar total mass. This leads to much bigger clumps,
with ${M\sub{T} \simeq 3\times 10^8}$~\msun{}. This value is typical for massive and gas rich galaxies \citep[see. e.g.][for in-depth discussion]{Genzel2008, Genzel2011, Guo2012, Tacconi2013, Tamburello2015}. {We have also attempted measuring masses of gas and stellar clumps in our simulations and found masses of similar order. We have plotted few most massive clumps on \autoref{fig:maps_agn}, measuring the mass in the radius of $4\Delta x\sub{min}\approx320\unit{pc}.$}
In conclusion, a sink particle with mass ${M\sub{sink} \le M\sub{T}}$ will have its trajectory easily disrupted by clumps in the disc. Larger sink masses, on the other hand, will result in a much more stable orbital evolution (see below).
	
In order for the sink particle to reach (or remain in) the nuclear region of the galaxy, we need to estimate the dynamical friction timescale as introduced by \cite{Chandrasekhar1943}.
Although the original formula was derived for a collisionless fluid (dark matter and stars), a very similar formula can be used to compute the dynamical friction on the gas \citep{Ostriker1999}.
For the gas drag, a correction factor must be introduced, compared to the original collisionless case, but only for transonic relative velocities. 
For a SMBH with a typical orbital velocity of 200~km/s, the drag force is likely to be in the strong supersonic regime, for which no correction is required.

Using Chandrasekhar's formula, we compute the dynamical friction timescale $t\sub{df}$ \citep[e.g. Eq. 8.12 of][]{Binney2008}
\begin{equation}
t\sub{df} = \frac{1.65}{\ln \Lambda}\frac{R\sub{orb}^2\sigma}{GM\sub{BH}}
\end{equation}
where the Coulomb logarithm is given by
\begin{equation}
\ln \Lambda =\ln\left(\frac{R\sub{gal}V\sub{orb}^2}{GM\sub{BH}}\right)
\label{eq:coulomb_log}
\end{equation}
$R\sub{orb}$ and $V\sub{orb}$ are the orbital radius and orbital velocity of the SMBH. 
Assuming that the velocity dispersion of the collisionless components $\sigma \simeq V\sub{gal}$,
the orbital radius and velocity of the SMBH to be of the order of the galaxy radius ${R\sub{gal} \simeq 5}$~kpc 
and circular velocity ${V\sub{gal} \simeq 200\,\mathrm{km\,s}^{-1}}$, 
and finally using $\ln \Lambda \simeq 6.9$ as a typical value for our purposes, 
we find the dynamical friction timescale to be
\begin{equation}
t\sub{df} \simeq 2.7 \mathrm{\,Gyr} \frac{10^8 \mathrm{\,M}_\odot}{M\sub{BH}}.
\end{equation}
Only SMBH with masses greater than $10^9$~\msun{} will be able to decay quickly enough to the centre of the galaxy, 
as they will have an orbital decay rate comparable or faster than their rotation rate. It is interesting to see that the Toomre mass
and the critical dynamical friction mass are both comparable to $10^9$~\msun{} in high-redshift Milky Way analogues \citep[see also][]{Bournaud2014}.
 
In a previous section, we have seen that a seed mass lower than $M\sub{Jeans}$ results in an artificially low, subsonic accretion rate. 
We see now that a seed mass lower than $M\sub{T}$ results in the sink particle being scattered out of the nuclear region by large gas clumps. 
Similarly, a large seed mass with a dynamical friction time scale comparable to (or shorter than) the orbital time $t\sub{orb} \simeq 200$~Myr will
help maintaining the sink particle within the nuclear region. 

In summary, large initial seed masses ($10^8$ and $10^9$ \msun{}) have a larger accretion rate, 
as ${\dot{M}\sub{Bondi}\propto M\sub{BH}^2}$, so they can grow fast, at their Eddington-limited rate, and become quickly less sensitive to orbital perturbations. 
Furthermore, $M\sub{T}$ is comparable to $M\sub{seed}$, thus sink particles do not suffer from encounters with larger mass perturbers. 
Also, its dynamical friction timescale is relatively short, helping the SMBH to remain in the centre. 
		
\subsection{AGN feedback-limited growth with supernovae feedback}\label{ssec:results_sn_agn}

\begin{figure*}
\begin{subfigure}{0.45\textwidth}
\centering
\includegraphics[width=\columnwidth]{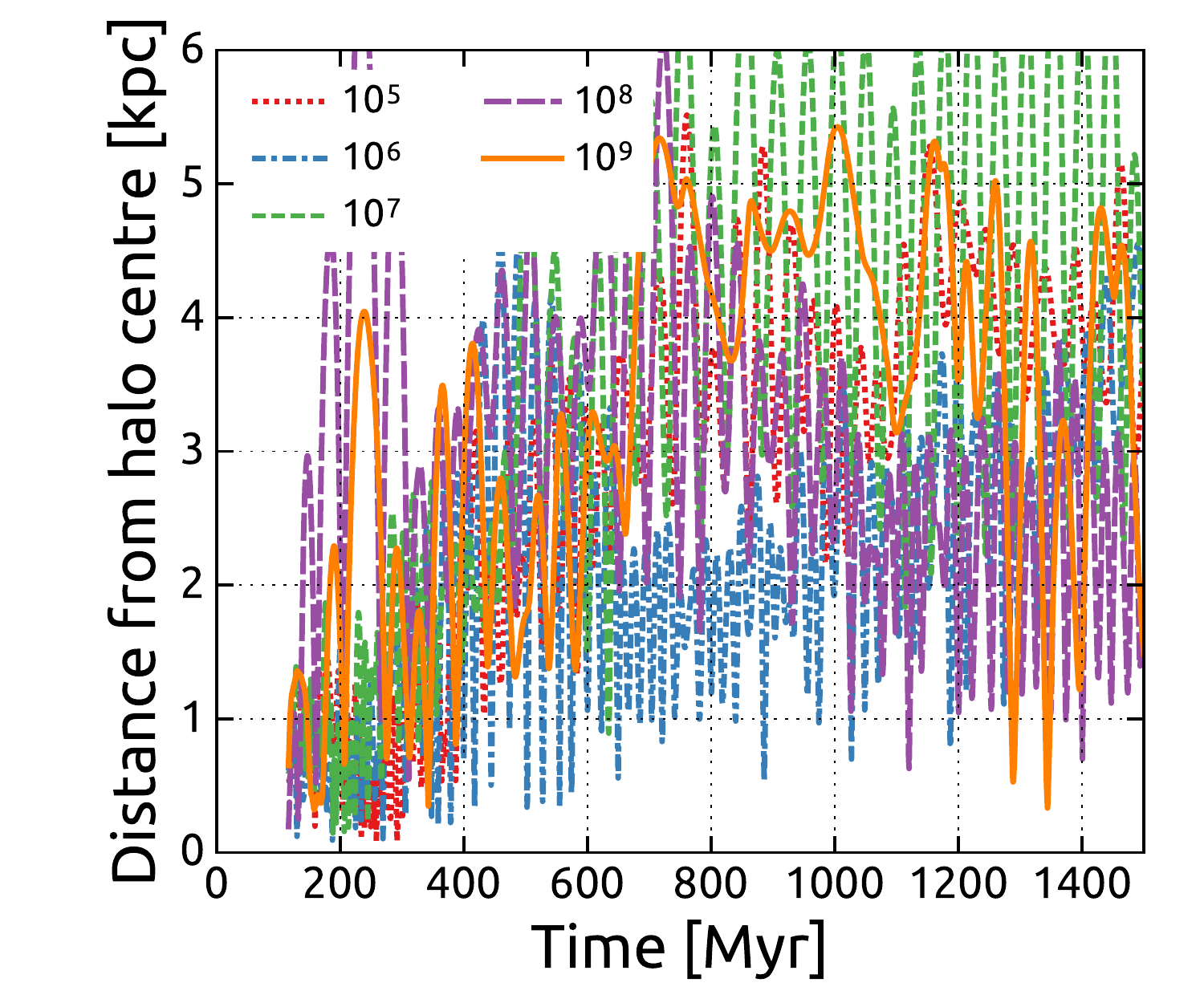}
\caption{Distance between the sink and the centre of the halo}\label{fig:dist_cent_sn_agn}
\end{subfigure}
\qquad
\begin{subfigure}{0.45\textwidth}
\centering
\includegraphics[width=\columnwidth]{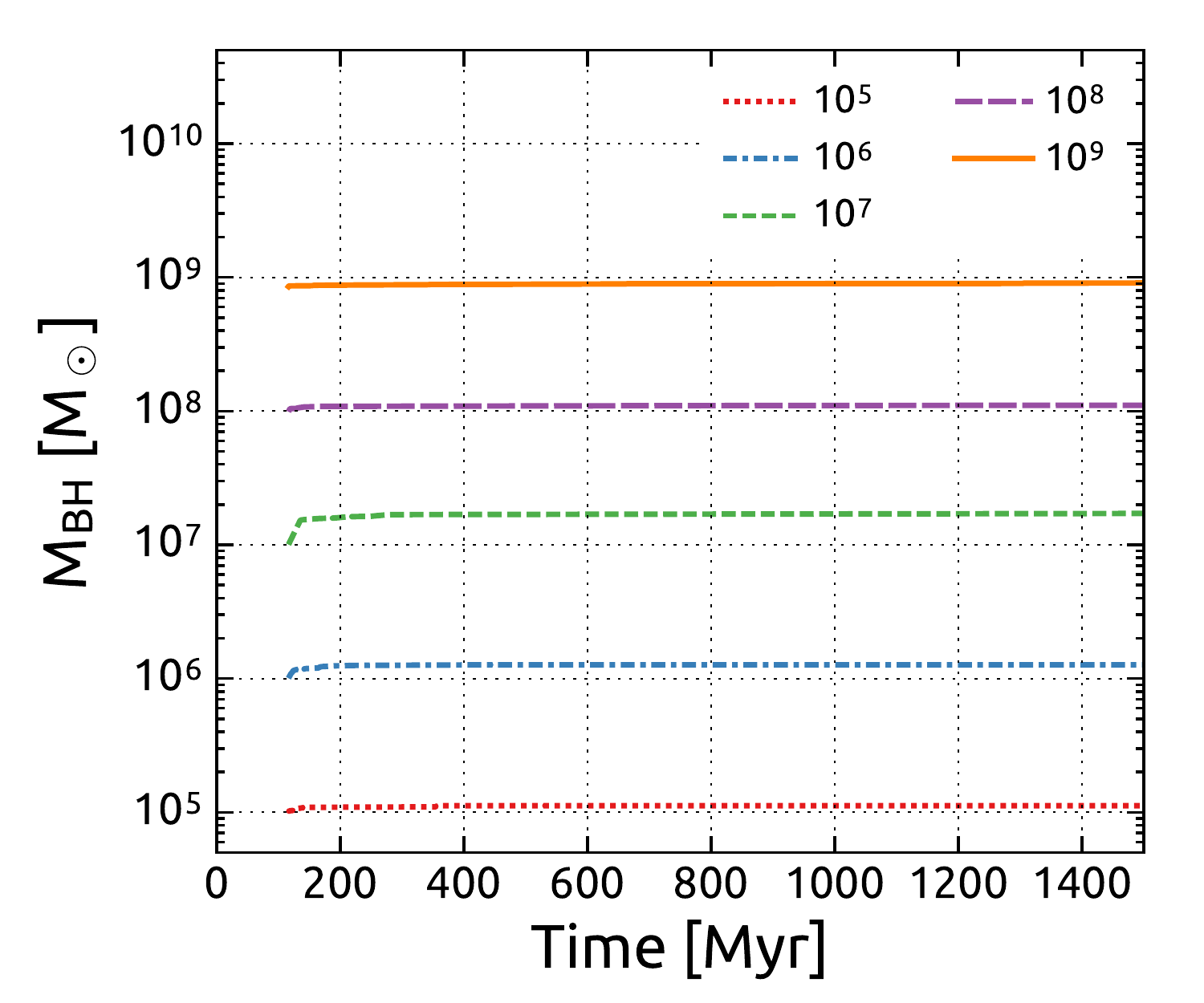}
\subcaption{SMBH mass growth}\label{fig:mass_evo_sn_agn}
\end{subfigure}
\caption{Evolution of distance to the centre of halo and sink mass for the runs with \emph{both} SN and AGN feedbacks for five different seed masses: {$10^5$~\msun{} - red (dotted), $10^6$~\msun{}  - blue (dash-dotted), $10^7$~\msun{}  - green (short dashes), $10^8$~\msun{}  - purple (long dashes), and $10^9$~\msun{}  - orange (solid). Grey band on the right panel shows predicted SMBH mass based on the halo escape velocity (cf. \autoref{eq:msinkcrit_vesc}).}}
\label{fig:hsc_sn_agn}
\end{figure*}

We now combine supernova and AGN feedback, repeating the same numerical experiments.
As before, the low and intermediate seed masses \mseed{5}, \mseed{6} and \mseed{7} do not really grow, as can be seen in \autoref{fig:hsc_sn_agn} for the {red (dotted), blue (dash-dotted) and green (short dashes)} lines,
and as it was already the case for our supernova-only feedback model.
The large seed mass, on the other hand, are already too close or even larger than their maximum, self-regulated SMBH mass, as it was already the case for our AGN-only feedback model. 
So even these large seed masses do not favour a fast growth of the sink particles, which are continuously perturbed by clumps with mass comparable or smaller than the Toomre mass.
Moreover, since the sink mass is not growing much beyond $10^9$~\msun{}, the dynamical friction time scale remains longer or comparable to the orbital time scale and the sink particles
keep moving around with eccentric orbits and large pericentre radii (see \autoref{fig:hsc_sn_agn} with violet and orange lines; also \autoref{fig:maps_agn}, left column). 

SN and AGN feedbacks work hand in hand to completely prevent SMBH growth in this gas rich, highly turbulent and clumpy environment. 
We argue that only SMBH already as massive as $10^{10}$~\msun{} can survive in the nuclear region of such a galactic  environment, 
because they resist the perturbations from clumps and because they have a short-enough dynamical friction time scale.
This conclusion is of course valid only if one considers that our two feedback models are realistic enough, which is of course highly speculative, since they rely on sub-grid physics.
These models are nevertheless quite state-of-the-art, and are required to explain the low star formation efficiency (for SN feedback) and to explain star formation quenching in massive galaxies (for AGN feedback). 

The fact that SMBH cannot grow at all (except the extremely massive ones) if one combines the two sources of feedback energy is therefore a fundamental problem 
in the theory of SMBH growth and co-evolution with galaxies. This also explains why many authors have to rely on artificial tricks to maintain the SMBH within the nuclear regions of galaxies,
especially when performing high-resolution simulations.

\subsection{Growth within a Nuclear Star Cluster}
\label{ssec:results_sn_nsc}
	
\begin{figure}
\centering
\includegraphics[width=\columnwidth]{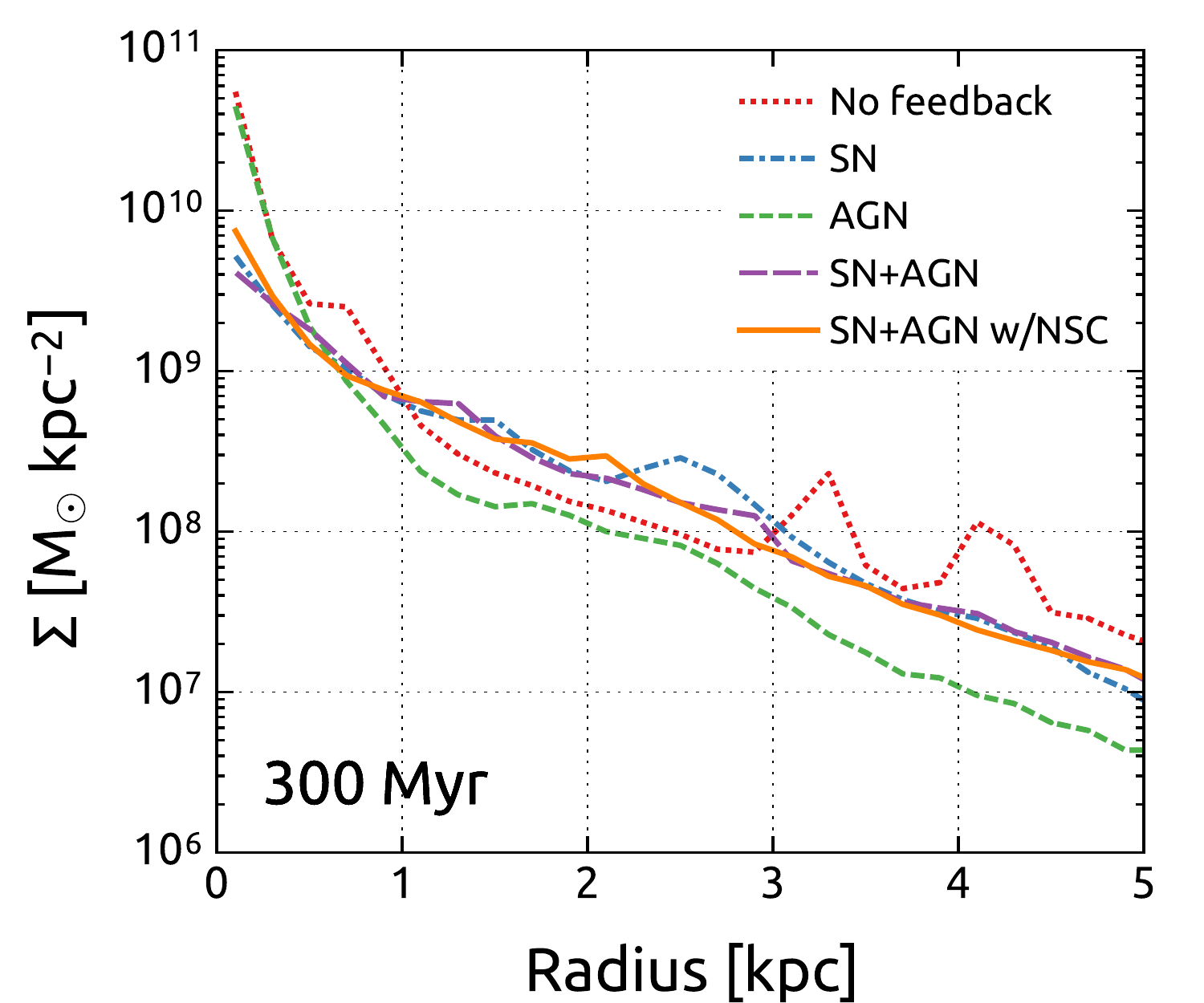}
\caption{Stellar density profile at 300 Myr for five different simulations: no feedback (red), AGN-only (blue), SN-only (green), SN+AGN (purple), and SN+AGN with NSC modelling (orange). All the profiles are centred with a shrinking sphere technique with respect to the total halo mass. Absence of SN feedback leads to creation of steep stellar profile and much more massive stellar bulge than with runs with SN feedback.}
\label{fig:prof_many}
\end{figure}

One of the key difference between the simulation with supernova-feedback and the simulations without supernova-feedback is the presence of a massive bulge,
or in other words, a massive nuclear concentration of stars (see \autoref{fig:prof_many}). 
Indeed, in the no supernova feedback cases, we do form massive clumps of gas and stars with masses of the order of (or smaller than) the Toomre mass, 
that appear as bumps in the stellar surface density profile in \autoref{fig:prof_many}. 
These perturbers do not seem to have an effect on the sink particle in the nuclear region (see \autoref{fig:dist_cent_smbh} and \autoref{fig:dist_cent_agn}), 
even for the small seed masses. The reason lies in the deep potential well provided by the stellar bulge hosting the sink particle. The mass of the bulge appears as
large enough to resist the external perturbation and to promote efficient migration towards the centre, using the same arguments as before. 
It has been argued that the SMBH-bulge co-evolution can be robustly established through observed scaling relations, 
which is not necessarily the case for the SMBH-galaxy co-evolution \citep{Magorrian1998, Haring2004, Kormendy2013}. 

In observed galaxies, we do see massive and isolated SMBH in the nuclear region (like in Andromeda) but also smaller SMBH without a massive bugle to host them (like in the Milky Way). 
Good candidates for hosting and protecting embedded SMBH in bulge-less galaxies are nuclear star clusters (NSC). NSC are interesting candidates for a co-evolution scenario with SMBH
in many aspects. First, one of the plausible SMBH formation scenarios advocates for the seed to be born within a dense star cluster {\citep[e.g.][]{Kochanek1987, PortegiesZwart1999, Davies2011, Stone2017}}.
Second, NSC are indeed massive enough to survive the perturbations from gas clumps in the host galaxy. Third, NSC are particularly compact (between 1 and 10 pc in size), 
so they can can trap efficiently their host SMBH within their deep potential well. 

The formation of NSC is unfortunately not well understood. 
In our simulations, the supernova feedback model completely prevent the formation of large and dense star clusters, and our spatial resolution won't allow the survival of parsec-scale objects like NSC anyway. 
In order to explore this idea, we have implemented a simple subgrid model of a SMBH evolving within a NSC. 
In our prescription, the sink particle now represents both the NSC and the SMBH. 
The seed mass is chosen as before for the SMBH, and set to zero for the accompanying star cluster. 
The Bondi rate is computed using the total sink mass (SMBH plus NSC), and is distributed to each component assuming that the NSC mass grows at a rate 100 times larger than that of SMBH. The Eddington limit is applied only to the SMBH growth rate.

\begin{figure*}
\begin{subfigure}{0.45\textwidth}
\centering
\includegraphics[width=\columnwidth]{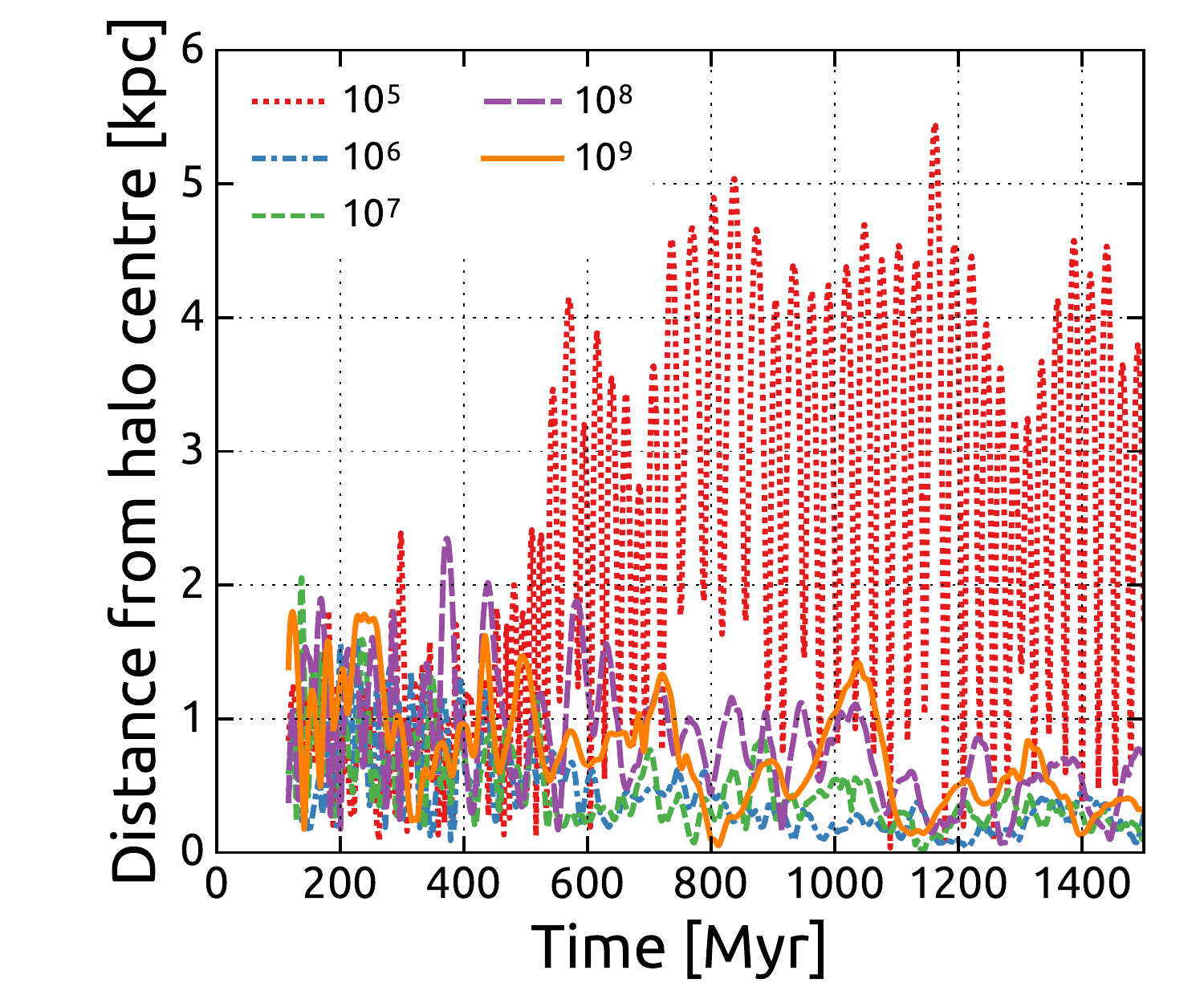}
\caption{Distance between the sink and the centre of the halo}\label{fig:dist_cent_sn_nsc}
\end{subfigure}
\qquad
\begin{subfigure}{0.45\textwidth}
\centering
\includegraphics[width=\columnwidth]{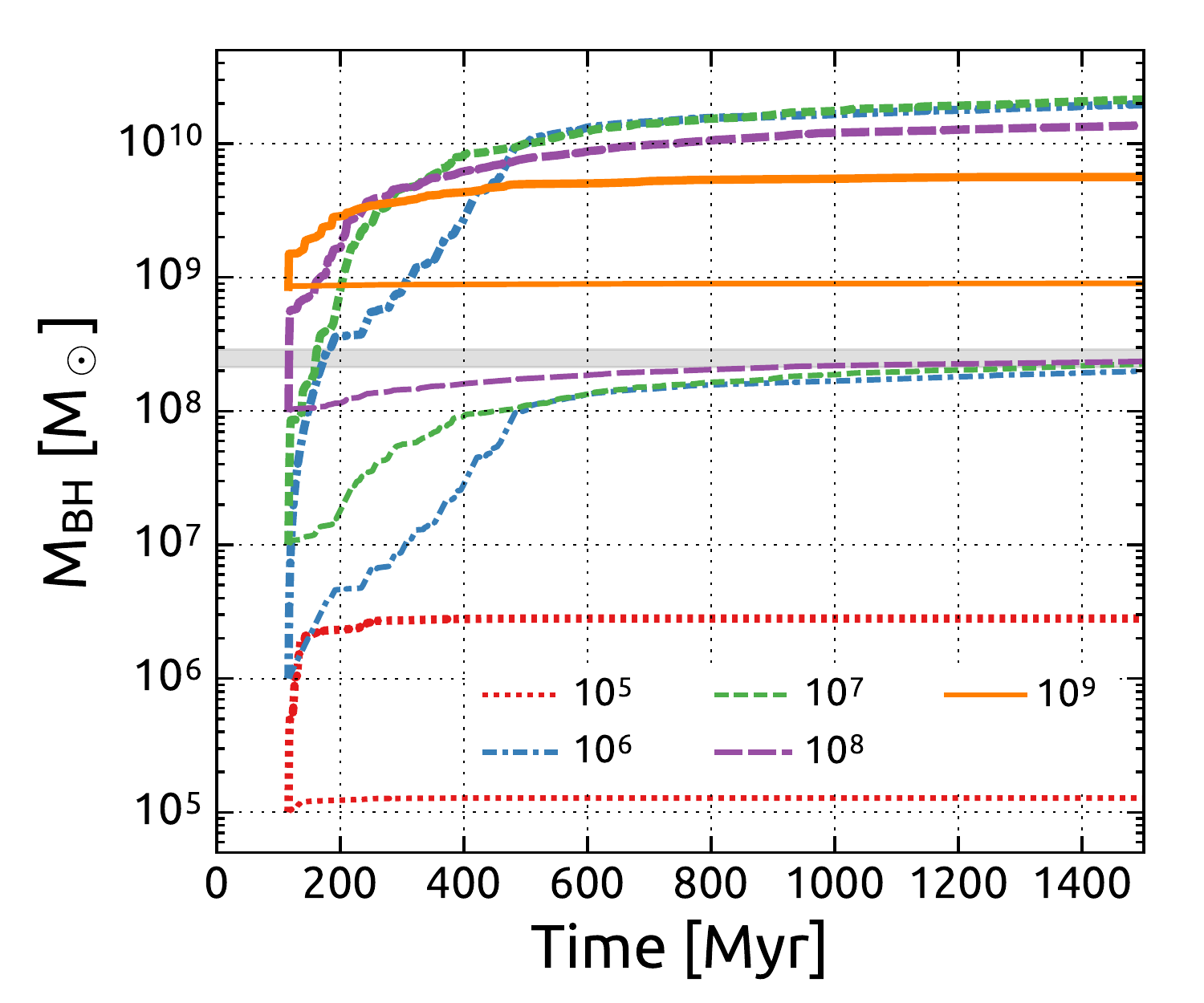}
\subcaption{SMBH (solid) and sink (NSC+SMBH; dashed) mass growth}\label{fig:mass_evo_sn_nsc}
\end{subfigure}
\caption{Evolution of distance to the centre of halo and sink mass for the runs with AGN feedback and NSC for five different seed masses: {$10^5$~\msun{} - red (dotted), $10^6$~\msun{}  - blue (dash-dotted), $10^7$~\msun{}  - green (short dashes), $10^8$~\msun{}  - purple (long dashes), and $10^9$~\msun{}  - orange (solid). Grey band on the right panel shows predicted SMBH mass based on the halo escape velocity (cf. \autoref{eq:msinkcrit_vesc}).}}
\label{fig:hsc_sn_nsc}
\end{figure*}

This model is arguably simplistic, and could be improved in many ways, for example by including more star cluster formation physics. 
Our goal here is to test this idea by analysing the dynamics of the resulting SMBH/NSC co-evolving system.
In \autoref{fig:dist_cent_sn_nsc} and \autoref{fig:mass_evo_sn_nsc}, 
we show our results for the combined SN and AGN feedback scenario with a NSC and for five different SMBH seed masses, as in the previous sections. 
The right panel now shows with a {thick} line the evolution of the NSC mass, while the SMBH mass is shown {with thin lines as before. Grey band shows analytical prediction from \autoref{eq:msinkcrit_vesc} for $v\sub{esc}=680\unit{km/s}$ as measured in the centre of the halo. It can be seen that final mass of the SMBH strongly depends on properties of the host halo. Slow and fractional growth after self-regulation should be attributed to loss of angular momentum by the gas in the galactic disc, as seen also in \autoref{ssec:results_smbh}.}

Compared to the similar scenario without NSC, one clearly sees that the sink particle remains now in the central kiloparsec (\autoref{fig:maps_agn}, right column), 
with the exception of the very low seed mass case, which still violates our Jeans mass condition.     

\begin{figure*}
\centering
\includegraphics[width=\textwidth]{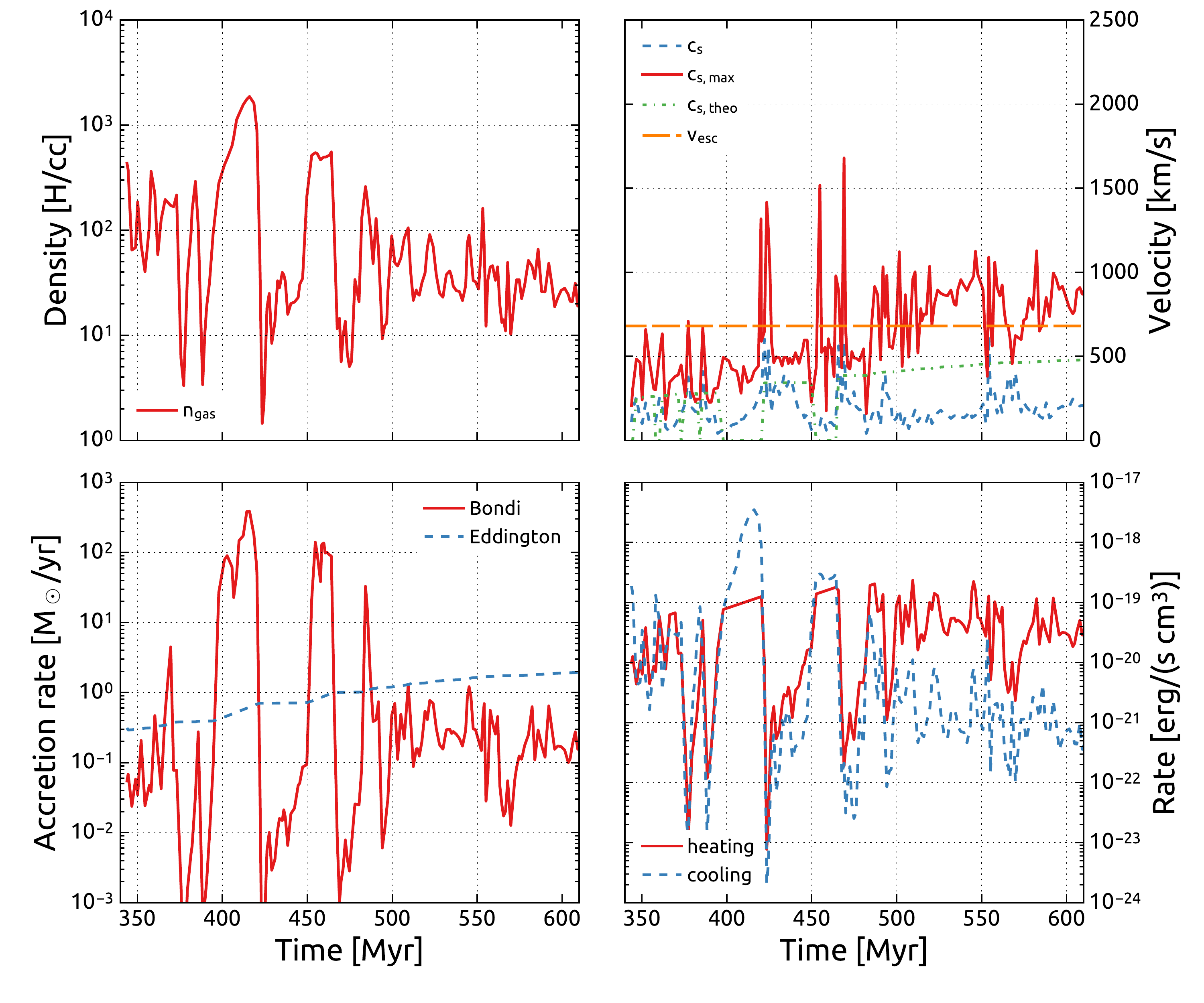}
\caption{Time evolution of 1) average gas density within the sink sphere (\emph{top left}); 2) average, mass-weighted, sound speed {(blue, short dashes) and maximum sound speed (red, solid) within the sink sphere (\emph{top right}), we have also represented our simple theoretical model (Eq.~\ref{eq:csad1} and \ref{eq:csad}) (green, dot-dashed) compared to the escape velocity from halo's centre (orange, long dashes); 3) Bondi (red, solid) and Eddington (blue, dashed) accretion rates (\emph{bottom left}) and 4)  average heating (red, solid) and cooling (blue, dashed) rates within the sink sphere (\emph{bottom right}) for simulation with SN and AGN feedbacks and NSC modelling and \mseed{6}.}}
\label{fig:vs_sn_nsc}
\end{figure*}

For SMBH with initial mass between $10^6$ and $10^8$ \msun{} initial growth is not Eddington-limited, but appears to be regulated by SN feedback, as they accrete at a sub-Eddington rate (\autoref{fig:vs_sn_nsc}).
The corresponding NSC mass is much larger, close to $10^{10}$~\msun{}, explaining why the combined NSC/SMBH system can survive interactions with clumps and remain in the centre.
Interestingly, the final NSC mass seems to depend on the initial SMBH seed mass. We explain this effect by the earlier termination of NSC growth due to AGN feedback. 
In our scenario, the NSC mass is assembled by fast, SN-regulated Bondi accretion, but is regulated ultimately by AGN feedback. 
The largest seed mass (\mseed{9}) has already reached the self-regulated mass scale and therefore does not grow at all, while its companion NSC can only grow its mass by a factor of 5.

\subsection{Resolution effects}\label{ssec:res_effects}

In order to determine to what extent the evolution of our SMBH mass is sensitive to resolution effects, 
we have re-run our various simulations with seed mass \mseed{6} and with a better spatial resolution with $\ell \sub{max}=15$ and a better mass resolution
with $m\sub{res} \simeq 2 \times 10^4 M\sub{\odot}$.

A complication that arises with this exercise is that massive gas clumps will form at different times in the fragmenting disks with different resolutions.
To avoid artificial differences due to stochastic effects, we have run the simulations with our highest resolution first, and then introduce the seed sink particle
at exactly the same time in the most massive clump of the lowest resolution afterwards. 

The AGN-feedback-only case at the different resolutions appears very similar in term of mass growth (basically Eddington-limited) but the final SMBH mass is larger in the high resolution run by a factor of 2. Using \autoref{eq:msinkcrit_cool}, we see that the critical SMBH mass for which cooling is balanced by heating, is proportional to the volume of the sink sphere, so that it should be reduced by a factor of 8 in the high resolution run, but is also proportional to the square of the gas density within the sink sphere, which happens to be 4 times larger in the high resolution case with $n\sub{H} \simeq 3000$~H/cc than in the low resolution case with $n\sub{H}\simeq 750$~H/cc, so that the critical mass should be increased by a factor of 16. Overall, as observed in the high resolution run, the final sink mass is larger by a factor of 2 when compared to the low resolution case.
The density in the sink sphere appears to be the critical parameter that controls the final sink mass, because of the delicate balance between heating and cooling. 
Before the SMBH mass is large enough to overcome the effect of cooling, nothing can prevent the collapse of the cold gas in the nuclear region (we do not include SN feedback yet) and the gas density can grow, up to a maximum value set by the adopted resolution. In conclusion, we argue that in this case (AGN-feedback only) {the final, maximum mass is set by the SMBH's ability to overcome cooling with heating}, and not its ability to heat the gas at (or above) the escape velocity of the halo.

\begin{figure*}
\begin{subfigure}{\textwidth}
\centering
\includegraphics[width=\columnwidth]{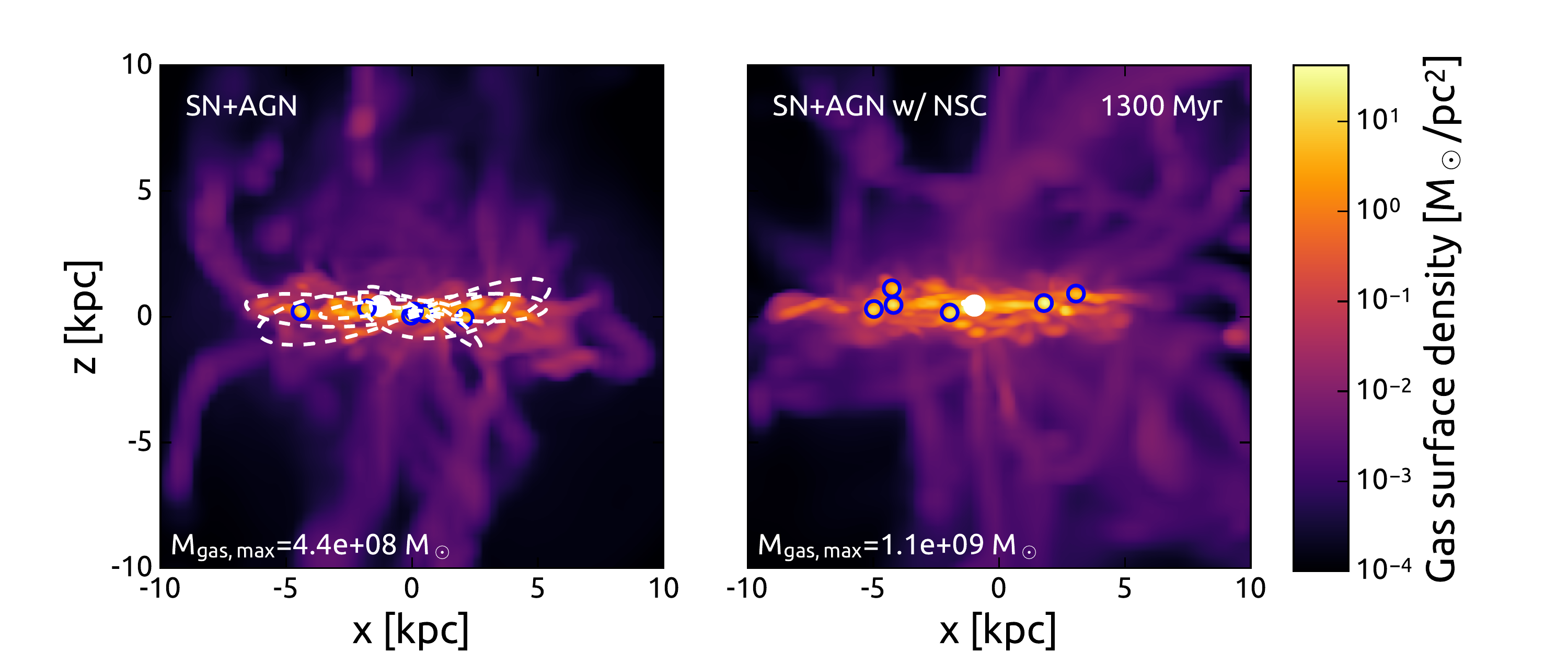}
\caption{Edge-on {volume-weighed} projection of gas surface density} \label{fig:gas_map}
\end{subfigure}
\\
\begin{subfigure}{\textwidth}
\centering
\includegraphics[width=\columnwidth]{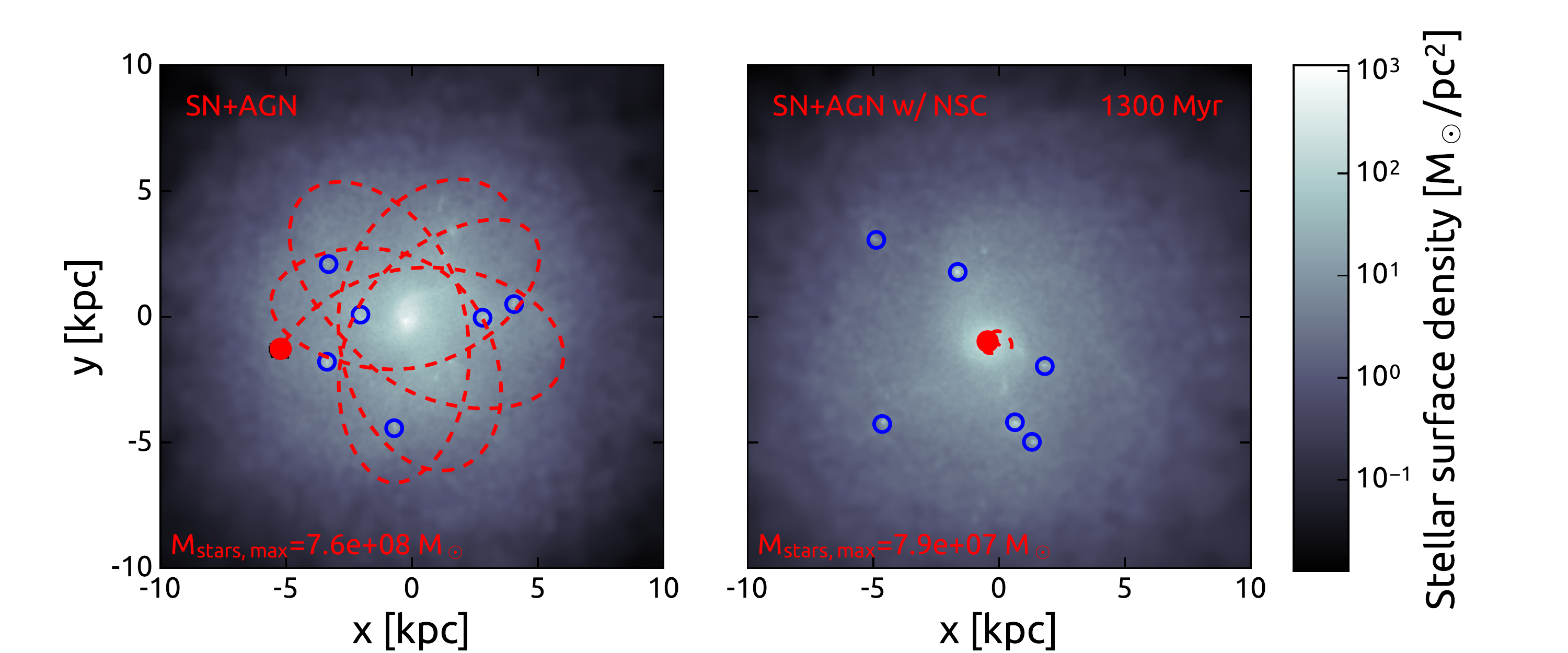}
\caption{Face-on {volume-weighed} projection of stellar surface density}\label{fig:stars_map}
\end{subfigure}
\caption{{Volume-weighed} projections of gas and stellar surface densities at 1300 Myr for the lower resolution run \emph{without} NSC (left column) and \emph{with} NSC (right column). The position of the sink is marked with a dot, while the dashed line marks past 100 Myr of sink's orbit; $M\sub{seed}=10^7$ \msun{} for all runs. Blue circles mark positions of few most massive clumps ($r\sub{clump}=320\unit{pc}$). (Movies showing dynamical evolution in these two runs can be found at \url{https://youtu.be/uFcV0u_MFOs} (without NSC) and \url{https://youtu.be/U0yNnAPTnmA} (with NSC).)}\label{fig:maps_agn}
\end{figure*}

When we include SN feedback (but without the NSC), the high resolution simulation is identical to the low resolution one, with the sink particle quickly moving out of the nuclear region on eccentric orbits and not growing at all (\autoref{fig:maps_agn}, left column, at lower resolution). The high resolution simulation shows SMBH orbits with systematically smaller apocentres, which is consistent with a slightly larger dynamical friction owing to the larger Coulomb logarithm due to the higher spatial resolution. Note that in the other two cases (AGN feedback only or NSC) the sink particles always remain in the central kpc (\autoref{fig:maps_agn}, right column, at lower resolution), independently on the adopted resolution.

\begin{figure*}
\begin{subfigure}{0.45\textwidth}
\centering
\includegraphics[width=\columnwidth]{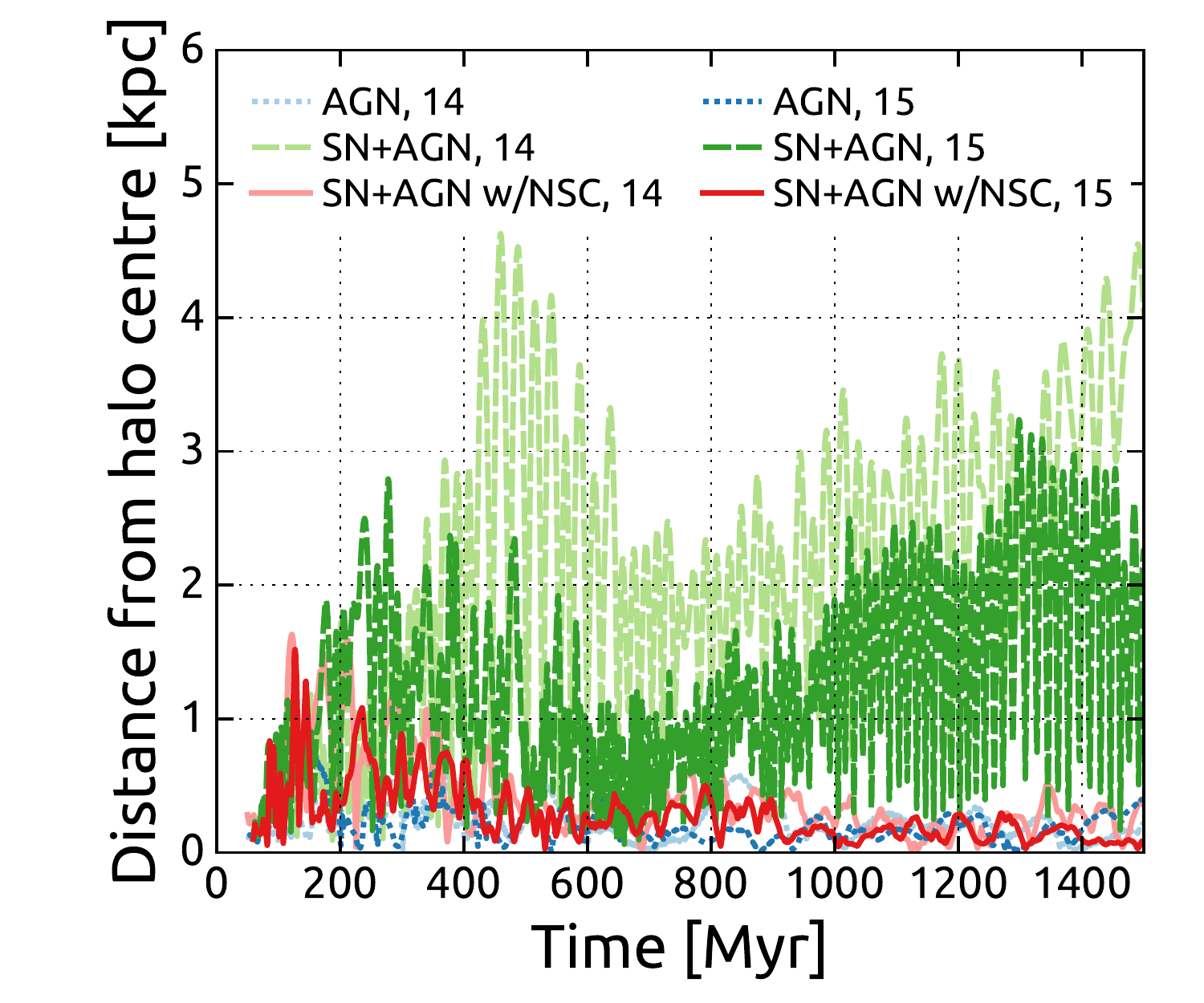}
\caption{Distance between the sink and the centre of the halo} \label{fig:ll_distance}
\end{subfigure}
\qquad
\begin{subfigure}{0.45\textwidth}
\centering
\includegraphics[width=\columnwidth]{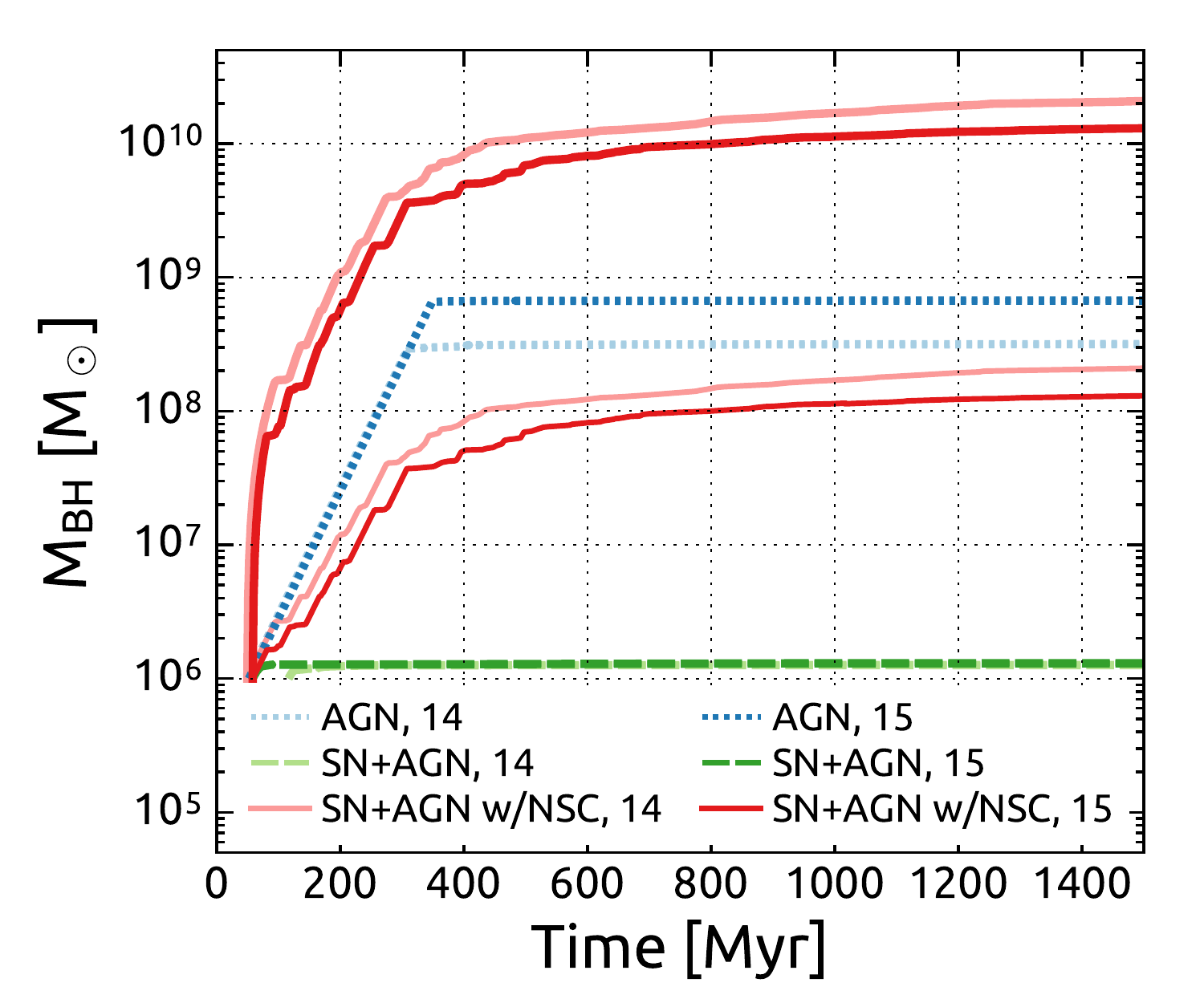}
\caption{{SMBH (thin) and sink (NSC+SMBH; thick) mass growth}}\label{fig:ll_msink}
\end{subfigure}
	\caption{Distance between halo centre and the SMBH and mass evolution of SMBH for two resolutions {(light for $l\sub{max}=14$ and dark for $l\sub{max}=15$) - runs with AGN feedback only (blue, dotted), SN+AGN (green, dashed) and SN+AGN with NSC (red, solid); dashed lines mark SMBH+NSC masses}. Accompanying videos can be found at: \url{https://youtu.be/1ECgXkrGv3U} (AGN), \url{https://youtu.be/DSeT_5ErJDY} (SN+AGN) and \url{https://youtu.be/SmMMdO4OL7s} (SN+AGN w/ NSC).}\label{fig:ll_comp}
\end{figure*}

When we finally include our NSC model, with both AGN feedback and SN feedback, the final sink mass appears to depend much less on resolution than the AGN-only case.
Because of SN feedback, we have now a succession of intense star forming events, where SN explosions blow the gas out of the sink sphere, hence reducing the gas density and helping AGN heating win over gas cooling, followed by quiescent phases when that gas can fall back again, so that cooling can win over heating, and the SMBH can grow fast (\autoref{fig:vs_sn_nsc} for lower resolution run).
Overall, the time-averaged density within the sink sphere is controlled (and significantly reduced) by SN feedback. The critical mass set by the balance between cooling and heating is therefore reduced, especially when the gas is completely gone. We can assume we are mostly in the adiabatic regime, and what matters in this regime is the ability of the SMBH to heat the gas at (or above) the escape velocity of the halo.
Using \autoref{eq:csad1}, we see that the final sink mass should be proportional to the cubic root of the adopted resolution, 
which is exactly what we observe in \autoref{fig:ll_comp}, where the final sink mass in the high resolution run is slightly smaller, 
but comparable to the final sink mass in the low resolution run. 
 
\section{Discussion}
\label{sec:discussion}

The evolution of SMBHs has been studied in two different regimes: 
1) the merging of binary black holes with sub-AU spatial resolution simulations 
\citep[][to name a few]{Chapon2013, Fiacconi2013, Roskar2015, SouzaLima2016} and 
2) the co-evolution of AGN and their host galaxies in cosmological simulations with spatial resolution of hundreds of parsecs at best 
\citep{Booth2009, Kim2011, Choi2012, Costa2014, Dubois2015, Tremmel2015}. 
The former is dedicated to the detailed study of the dynamics of binary black holes within a nuclear gas disk at very high spatial resolution, 
while the latter often discards the dynamical evolution of the SMBHs, as many of the relevant scales are not resolved. 
In this work, we attempt to bridge the gap between those two different approaches, focusing on the detailed dynamics of the central SMBH while retaining the large scale galactic evolution.

Recent work by \cite{Fiacconi2013, Roskar2015} and \cite{SouzaLima2016} have showed that binary black holes can be scattered outside of the nuclear disc, 
if physical processes like SN feedback and gas cooling are present. The former produce outflows, which rarify the medium and  thus reduce the effect of dynamical friction, 
while the latter leads to the formation of gas clumps that can scatter the SMBH out of the disc plane. 
This is in complete agreement with what we have obtained in this paper on a larger scale and over a longer time scale.

Modelling the precise dynamics of SMBHs in cosmological simulations has not been the priority of galaxy formation simulators in the past decade. 
There is no consensus on the SMBH formation scenario and on their initial seeding environment. Very often, although AGN feedback is described at length,
very little has been said about the possibly complex dynamics of SMBH within their host galaxies \citep[see for example][]{Kim2011, Choi2012}. 
\cite{Gabor2013} have reported in their simulations of high-redshift galaxies that the central SMBH is scattered by massive clouds. They decided to add an artificial acceleration towards the centre of mass of the stellar component to maintain the SMBH in the nuclear region \citep[see also][]{Okamoto2008}. 
A similar effect has been observed by \cite{Bellovary2010}, but they did not correct for it. 
They obtained many SMBHs with orbits from 10 to 100 kpc from the centre of a halo. 
They argued that these large, eccentric orbits are physical, as the dynamical friction timescales of the wandering SMBHs are longer than age of the Universe.
They also found that low mass seeds grow on average only by 2\%. 

In \cite{Debuhr2011}, the authors used a different methodology by assigning a higher dynamical mass to their SMBH, set to 100 times the black hole mass, 
justifying it as a way to avoid ``Brownian'' motion and highly eccentric SMBH orbits. A similar approach was adopted in the simulations of \cite{Gabor2016}, 
but they used a constant dynamical mass of $10^9\,\mathrm{M}\sub{\odot}$. 
In \cite{Costa2014}, the authors followed the extreme strategy of entirely abandoning the dynamical evolution of the SMBH by keeping it fixed at the centre of the halo. 
Most recently, \cite{Sijacki2015} and \cite{Schaye2015} opted for repositioning the SMBH to the minimum of the gravitational potential at each time step, 
following a recipe similar to  \cite{Springel2005} and \cite{Booth2009}.  
Finally, \cite{Tremmel2015} followed a more physical approach, adding an explicit dynamical friction force to the SMBH acceleration, 
invoking their limited spatial resolution to correct for the underestimated Coulomb logarithm. 
All these different authors are trying to address the issue of the dynamics of the SMBH using various strategies, not always physically motivated.
In this work, we are attempting to address the same issue, using an additional physically motivated gas drag, 
or using an observationally motivated solution with the introduction of a companion NSC.

In Sections~\ref{ssec:results_sn_smbh} and \ref{ssec:results_sn_nsc}, we have seen that SN feedback can control or prevent the growth of the SMBH, mostly by triggering a complex dynamical evolution 
of the SMBH. A similar conclusion has been reached by \cite{Dubois2015}, for which SN feedback is responsible for regulating the SMBH growth in high-z halos. 
In the presence of strong SN feedback, cold gas is removed from the nuclear region. It is only once the bulge mass reaches $10^9$\msun{} that the gas flow can stabilise in the centre, 
so that the SMBH is able to accrete at the Eddington rate. They argue, that this is due a sudden increase of the escape velocity, which exceeds the velocity of SN-powered outflows. 
A similar argument has been used by \cite{Bower2017}. They used a simple analytical model to describe the central SMBH growth in the presence of hot, buoyantly rising, SN-driven outflow,
that limit the gas density in the immediate environment of the SMBH. 
Once the halo mass reaches $10^{12}$\msun{}, the SN bubbles are not buoyant anymore and the gas density can increase, leading to a fast SMBH growth.
Both arguments lead to a similar conclusion: in order for the SMBH to grow fast enough, it is required to meet the conditions to form a dense and massive enough central concentration, 
in the form of a stellar bulge or a dense, gas rich, nuclear region. 

In the simulations performed in this paper, we observe a similar effect. 
The scenarios for which a massive bulge can form, namely without feedback or with only AGN feedback are the only ones leading to a fast growth.
Using our efficient SN feedback recipe, we cannot form a large bulge, and our central SMBH does not grow. 
In our case, however, this is because of its erratic dynamical evolution. 
We argue in this paper that this is the complex dynamics of the SMBH that can prevent its fast growth, by reducing the Bondi accretion rate due to an increased relative velocity 
between the sink and the gas. Large, eccentric orbits are unavoidable, due to the combined effects of large mass perturbers and inefficient dynamical friction.
In order to stabilise the dynamics of the SMBH in the central kiloparsec, we propose another viable scenario, namely to attach to the SMBH a dense, compact and more massive NSC.

Our NSC hypothesis can be supported by local observations of SMBHs (see e.g. \citet{Graham2009} or the excellent review by \citet{Kormendy2013}). 
These show that SMBHs coexist with NSCs in the centres of galaxies, regardless of the type of the host \citep{Seth2008, Graham2009}. 
This hints towards a scenario in which SMBH coevolves with NSC. 
The protective environment of NSC is particularly important for the growing SMBH in the presence of massive perturbers in the galaxy. 
The perturbers can be either giant molecular clouds or stellar clusters. 
In the sample of \citet{Seth2008} the ratio $M_{\mathrm{SMBH}}/M_{\mathrm{NSC}}$ is typically between 0.01 and 1, which justifies our simple model for the NSC growth. 
\cite{Graham2009} also lists many galaxies with prominent nuclear component hosting a less massive SMBH. 

The nuclear region of the Milky Way (MW) hosts a relatively small SMBH with $M_{\mathrm{SMBH}} \simeq 3\times 10^6$~\msun{}, 
which is comparable to the typical mass of potentially perturbing GMCs, owing to the low gas fraction of the MW. 
The MW has no massive bulge, but hosts a NSC with mass ${3\times10^7}$~\msun{}, which can resist external perturbations. 
The corresponding dynamical friction time scale is quite large, 10~Gyr, but still comparable to the age of the Universe. 
Similarly, in the Circinus galaxy \citep{Maiolino1998}, a SMBH of mass $1.7 \times10^6$ ~\msun{} \citep{Gultekin2009} is believed to sit 
within a NSC of mass of $10^{7}$~\msun{} and located within the galactic bar, thus being well protected from perturbers. 

The Andromeda galaxy, on the other hand, hosts a central SMBH with mass of the order of $10^8$ \msun, 
which is about two orders of magnitude larger than the typical GMC mass in this galaxy \citep{Blitz2007, Rosolowsky2007}. 
So the Andromeda SMBH can resist alone external perturbations, and benefits from a relatively short, 3~Gyr, dynamical friction time scale.
Interestingly, the NSC in the Andromeda galaxy is four times less massive that its SMBH \citep{Kormendy2013}.

More massive galaxies (${M\sub{halo} \ge 10^{12}}$~\msun{}) are usually bugle-dominated or elliptical galaxies, 
and typically contain very massive SMBH with no sign of a companion NSC \citep{Graham2009}.

Smaller mass galaxies usually show SMBHs hosted by more massive NSC \citep{Graham2009}. 
For example, NGC~4395 is a small mass galaxy with $V\sub{max} \simeq 90$~km/s and total stellar mass within the galactic disk $M\sub{tot}\simeq10^9$~\msun{}. Mass estimates for dwarf galaxies are challenging, thus value of $M\sub{tot}$ quoted here for NGC4395 is at most factor of two larger (assuming $f\sub{gas}=0.5$), which would support our argument even more. 
A NSC of mass $1.4\times10^6$~\msun{} hosts one of the smallest mass SMBH (an Intermediate Mass Black Hole or IMBH) 
ever detected with $M\sub{SMBH}=3.2\times10^5$~\msun{} \citep{Seth2008,Graham2009,denBrok2015}.
These numbers are consistent with our scenario of SMBH and NSC co-evolution.
A similar galaxy, POX~52, contains a SMBH with mass also close to with ${10^5}$~\msun{} \citep{Barth2004,Thornton2008} and some indications of a companion NSC, 
although the evidence is not as clear as for the previous case \citep{Thornton2008}. 

The situation is somewhat more complicated at higher redshift ($z\simeq2$), at the peak of star formation, when galaxies are gas rich and fragmented into massive clumps \citep[see e.g. ][for discussion on importance of massive gas clumps for bulge and SMBH formation]{Elmegreen2008a, Elmegreen2008b}. There is no observational evidence that SMBHs are not hosted by giant NSCs in the early Universe, but see \cite{Schawinski2011} for a peculiar triple AGN galaxy.

Another argument in favour of our scenario is related to possible theories for the formation of NSC and SMBH/IMBH.
For the former, our simulations are consistent with the {\it in situ} formation scenario of \cite{Milosavljevic2004}, for which NSC form from collapsed gas in the nuclear region. 
For the latter, we invoke one possible scenario of IMBH formation based on runaway collisions of stars in a dense star cluster \citep{Kochanek1987, PortegiesZwart1999, Davies2011, Stone2017}, 
the star cluster being in our case the recently formed NSC. If the formation of both NSC and SMBH are related, then this could further support the idea of their subsequent co-evolution.
\cite{Gnedin2014} discuss in details this idea of co-formation of NSCs and SMBHs.
One serious caveat in this picture is that we do not observe any NSC associated to more massive SMBH in elliptical galaxies. This could be explained by the SMBH becoming massive enough 
to disperse the stars and evaporate the NSC \citep[e.g.][]{Merritt2009}.
	
\section{Summary and conclusions}\label{sec:summary}
	
In this work, we have presented and tested a new algorithm for SMBH modelling in the \ramses{} code. 
This method was designed on top of the previous work of \citet{Bleuler2015} in the context of star formation in molecular clouds. 
The new, upgraded sink particle algorithm is used here for the first time in the context of SMBH accretion and dynamical evolution, in conjunction with an AGN feedback model. 
We form SMBH seeds in massive gaseous clumps detected using the new clump finder \textsc{phew} \citep{Bleuler2015}. 
The SMBH growth is modelled via Eddington-limited Bondi accretion. 
Its dynamical evolution is treated carefully with a direct $N$-body integrator and including optionally a drag force due to exchange of momentum with the gas.

We have tested our new model within high-resolution simulations of an isolated, gas-rich cooling halo, whose properties appears very similar to high-z clumpy galaxies.
We have explored the effects of our new AGN feedback model on the growth and the orbital evolution of our central SMBH, in conjunction (or in competition) with an efficient model for SN feedback. 

In a control simulation without any feedback, we have shown that our sink particle remains trapped within a dense central bulge and accretes gas at the Eddington rate,
provided that the seed mass is larger than the minimum Jeans mass set by the mass resolution of our simulation. 
The final SMBH mass is regulated by gas accretion into the nuclear region, or in other words by starvation of the SMBH.

In the presence of AGN feedback only, we observe also the formation of a massive bulge and the SMBH grows quickly until it reaches a final mass self-regulated by AGN feedback.
We have developed a simple analytical model to support our findings and we argue that in absence of SN feedback, the final SMBH mass is equal to a critical mass for which AGN heating balances
gas cooling within the vicinity of the SMBH. When this happens, the SMBH can clear out the  gas from the nuclear region and stops growing.

In the presence of our efficient SN feedback model, we prevent the galaxy from forming a stellar bulge. 
As a consequence, the central SMBH is easily perturbed by massive gas clumps and quickly leave the nuclear region on highly eccentric orbits.
Due to a large relative velocity between the sink and the gas, its accretion rate drops and the SMBH stops growing. 
Only models with a high enough seed mass can grow fast enough to sustain external perturbation and maintain the SMBH in the centre.
 
Finally, using both feedback models together, we have shown that the central SMBH cannot grow at all, because of SN feedback for small seed mass, and because of AGN feedback for large seed mass.

To overcome this apparent dead end in the SMBH evolution in high-z, gas rich galaxies, and inspired by local observation of nuclear regions in nearby galaxies, 
we propose a new model in which SMBH are seeded and coevolve with a NSC. 
We have implemented a very simple model for the joint SMBH/NSC system, in which the NSC is allowed to grow fast enough to resist external perturbations
and to provide a short dynamical friction time scale, so that the sink particle can accrete mass efficiently and remain within the nuclear region. 
Interestingly, in this scenario, SN feedback is controlling the gas supply in the vicinity of the SMBH and the balance between gas cooling and AGN heating.
As a consequence, using our same analytical model, we show that the final SMBH mass is not determined by the balance between AGN heating and gas cooling anymore,
but instead by the balance between AGN heating and gravity, namely by comparing the gas sound speed to the halo escape velocity. 

In conclusion, we argue, using dynamical arguments, that the SMBH must remain in the nuclear region of the galaxy in order to grow fast enough.
This is possible only if the galaxy can grow a massive bulge or a dense NSC. We have shown that the latter scenario might be plausible, 
although our NSC formation and growth model could be improved significantly. 
We will show in a companion paper how this impact the star formation rate and the outflow properties in the parent galaxy.
		
\section*{Acknowledgements}

We would like to acknowledge stimulating conversations with Victor Debattista, Arif Babul, Massimo Dotti, Jillian Bellovary, Pedro R. Capelo and Davide Fiacconi. We thank the anonymous referee for helpful comments that improved this paper. Simulations performed for this work were executed on zBox4 at University of Zurich and on Piz Dora and Piz Daint at Swiss Supercomputing Center CSCS in Lugano.

\bibliographystyle{mnras}
\def\apj{ApJ}
\def\apjs{ApJS}
\def\apjl{ApJL}
\def\aj{AJ}
\def\mnras{MNRAS}
\def\aap{A\&A}
\def\nat{Nature}
\def\pasj{PASJ}
\def\prd{PRD}
\def\physrep{Physics Reports}
\def\jcap{JCAP}
\bibliography{dynamics}

\appendix
\section{Numerical implementation of the drag force}\label{sec:appendix}

In this Appendix we expand on our implementation of the accretion-related drag force.

We solve Eqs. (\ref{eq:drag1})-(\ref{eq:drag2}) to get
\begin{eqnarray*}
\vect{x\sub{sink}^{n+1}}\left(\tilde{M}+M\sub{acc}\right) & = &  \vect{x}\sub{COM} M\sub{acc}+\vect{x\sub{sink}^n}\tilde{M},\nonumber\\
\vect{x\sub{gas}^{n+1}}\left(\tilde{M}+M\sub{acc}\right) & = &  \vect{x}\sub{COM} M\sub{acc}+\vect{x\sub{gas}^n}\tilde{M},
\end{eqnarray*}
where $M\sub{acc}=\max{(0,\dot{M}\sub{Bondi}-\dot{M}\sub{acc})\upright{d}t}$ and
\begin{eqnarray}
\tilde{M} & = & \frac{M\sub{gas} M\sub{sink}}{M\sub{gas} + M\sub{sink}},\\
\vect{x}\sub{COM} & = & \frac{M\sub{gas} \vect{x\sub{gas}^n} + M\sub{sink} \vect{x\sub{sink}^n}}{M\sub{gas} + M\sub{sink}}.
\end{eqnarray}
This then leads to the change of sink position by
\begin{equation}
\Delta \vect{x\sub{sink}} = \frac{\tilde{M}M\sub{acc}}{\tilde{M}+M\sub{acc}}\left(\vect{x\sub{gas}^n}-\vect{x\sub{sink}^n}\right).
\end{equation}
The same can be written for momentum and one obtains
\begin{equation}
\vect{p\sub{drag}}= \frac{\tilde{M}M\sub{acc}}{\tilde{M}+M\sub{acc}}\left(\vect{v\sub{gas}^n}-\vect{v\sub{sink}^n}\right),\label{eq:p_drag}
\end{equation}
where the sink contribution is weighted by the mass of the gas.
\par The complete drag modelling requires modifying the state of the gas around the sink. This can be written as
\begin{equation}
\Delta \rho\sub{gas} = \left(M\sub{acc}\left.\right|\sub{M}-M\sub{virt}\left.\right|\sub{M}+M\sub{virt}\left.\right|\sub{V}\right)V^{-1},
\end{equation}
where $M\sub{virt}=(\dot{M}\sub{BH}-\dot{M}\sub{acc})\upright{d}t$ is a \textit{virtual accreted mass} and $\left.\right|\sub{x}$ denotes weighting with $x$ variable, here mass $M$ and volume $V$ of the gas around the sink particle. Thus, change of density of the unit gas cell is due to regular, Eddington-limited accretion and due to redistribution of the remaining, Bondi-accreted gas around the sink. The latter can be seen as a physical manifestation of the Eddington pressure. The momentum of a gas cell is modified by the momentum of the accreted material as well as by the momentum exchange with the sink - essentially \autoref{eq:p_drag} weighted by the fractional volume of the cell with respect to the volume of the sink accretion sphere. The energy state of the gas has to be modified to account for the  truly accreted material as well as for the \textit{decreted} specific internal energy $\varepsilon$. We can write
\begin{eqnarray}
\rho \frac{\upright{d}\varepsilon}{\upright{d}t}  & = & -p \nabla \cdot \vect{v} \\
\rho \frac{\upright{d}\vect{v}}{\upright{d}t} &  = & -\nabla \vect{p}+\vect{F}\sub{drag},
\end{eqnarray}
and with
\begin{equation}
E\sub{tot}=\frac{1}{2}\rho u^2+ \rho\varepsilon
\end{equation}
we can solve the above to obtain
\begin{equation}
\frac{\partial}{\partial t}\left(E\sub{tot}\right)+\nabla\cdot\left(\vect{v}E\sub{tot}+ \vect{v}p\right)=\vect{F}\sub{drag}\cdot\vect{v}.
\end{equation}

\begin{figure}
\centering
\includegraphics[width=\columnwidth]{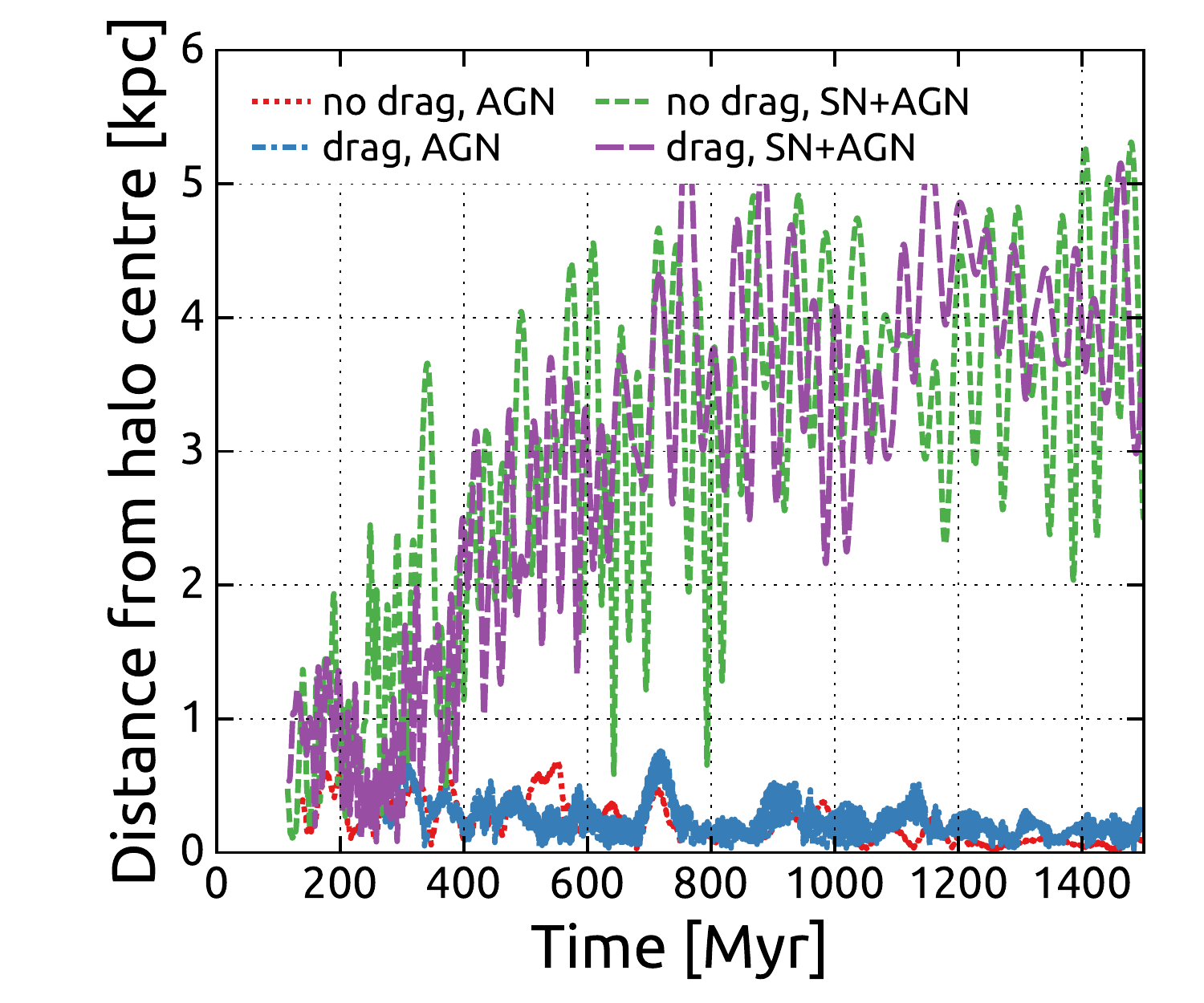}
\caption{Evolution of distance to the centre of halo for the runs with AGN feedback and without drag - red (dotted), AGN feedback and drag - blue (dash-dotted), SN+AGN feedbacks without drag - green (short dashes), SN+AGN feedbacks with drag purple (long dashes).}
\label{fig:drags}
\end{figure}

{We test our implementation in two different setups: 1) with AGN feedback only and 2) with both SN and AGN feedbacks present and show the results on \autoref{fig:drags}. It can be seen that modelling of the accretion drag does not help to lock the sink in the halo centre.}

{In the Eddington limited phases of accretion in simulations without SN feedback, additional drag is not needed, as gas hosts are long-lived and provide enough protection for the growing seed (see \autoref{ssec:results_smbh} and \ref{ssec:results_agn}). If SN feedback is enabled it is that process that regulates the fate of SMBH - here Eddington-limited accretion episodes are short at best (cf. \autoref{fig:vs_sn_nsc}), thus not influencing dynamical evolution of sink enough. Despite its apparent insignificance we do include the accretion drag prescription in all our runs, as we want to create the most favourable conditions for SMBH growth.}

\end{document}